\documentclass[final,3p,times,twocolumn]{elsarticle}
\usepackage{graphicx}
\usepackage{dcolumn}
\usepackage{amssymb,amsmath}
\usepackage{natbib}
\usepackage{subfigure}
\usepackage{epstopdf}
\usepackage{placeins}
\usepackage{citesort}
\usepackage{color}
\usepackage{amssymb}

\journal{European Journal of Mechanics, B/Fluids}

\begin{document}

\begin{frontmatter}



\title{Heat transport and flow structure in rotating Rayleigh-B\'enard convection}


\author{Richard J. A. M. Stevens$^{1}$, Herman Clercx$^{2,3}$ and Detlef Lohse$^{1}$ }

\address{$^{1}$Department of Physics, Mesa+ Institute,  and J.\ M.\ Burgers Centre for Fluid Dynamics, University of Twente, 7500 AE Enschede, The Netherlands\\
$^2$Department of Applied Mathematics, University of Twente, Enschede, The Netherlands\\
$^3$Department of Physics and J.M. Burgers Centre for Fluid Dynamics, Eindhoven University of Technology, P.O. Box 513, 5600 MB Eindhoven, The Netherlands}

\begin{abstract}
Here we summarize the results from our direct numerical simulations (DNS) and experimental measurements on rotating Rayleigh-B\'enard (RB) convection. Our experiments and simulations are performed in cylindrical samples with an aspect ratio $\Gamma$ varying from $1/2$ to $2$. Here $\Gamma=D/L$, where $D$ and $L$ are the diameter and height of the sample, respectively. When the rotation rate is increased, while a fixed temperature difference between the hot bottom and cold top plate is maintained, a sharp increase in the heat transfer is observed before the heat transfer drops drastically at stronger rotation rates. Here we focus on the question of how the heat transfer enhancement with respect to the non-rotating case depends on the Rayleigh number $Ra$, the Prandtl number $Pr$, and the rotation rate, indicated by the Rossby number $Ro$. Special attention will be given to the influence of the aspect ratio on the rotation rate that is required to get heat transport enhancement. In addition, we will discuss the relation between the heat transfer and the large scale flow structures that are formed in the different regimes of rotating RB convection and how the different regimes can be identified in experiments and simulations.
\end{abstract}


\end{frontmatter}


Rayleigh-B\'enard (RB) convection, i.e.\ the flow of a fluid heated from below and cooled from above, is the classical system to study thermally driven turbulence in confined space \cite{ahl09,loh10}. Buoyancy-driven flows play a role in many natural phenomena and technological applications. In many cases the fluid flow is also affected by rotation, for example, in geophysical flows, astrophysical flows, and flows in technology \cite{joh98}. On Earth, many large-scale fluid motions are driven by temperature-induced buoyancy, while the length scales of these phenomena are large enough to be influenced by the Earth's rotation. Key examples include the convection in the atmosphere \cite{har01} and oceans \cite{mar99}, including the global thermohaline circulation \cite{rah00}. These natural phenomena are crucial for the Earth's climate. Rotating thermal convection also plays a significant  role in the spontaneous reversals of the Earth's magnetic field \cite{has99}. Rotating RB convection is the relevant model to study the fundamental influence of rotation on thermal convection in order to better understand the basic physics of these problems.

In this paper we discuss the recent progress that has been made in the field of rotating RB convection. First we discuss the dimensionless parameters that are used to describe the system. Subsequently, we give an overview of the parameter regimes in which the heat transport in rotating RB is measured in experiments and direct numerical simulations (DNSs). This will be followed by a description of the characteristics of the Nusselt number measurements and a description of the flow structures in the different regimes of rotating RB. Finally, we address how the different turbulent states are identified in experiments and simulations by flow visualization, detection of vortices, and from sidewall temperature measurements.

\section{Rotating RB convection}
When a classical RB sample is rotated around its center axis, it is called rotating RB convection. For not too large temperature gradients, this system can be described with the Boussinesq approximation
\begin{eqnarray}\label{Intro_NS_rot}
\frac{\partial \textbf{u}}{\partial t} + \textbf{u} \cdot \nabla \textbf{u} + 2 \boldsymbol \Omega \times \textbf{u} &=& -\nabla p + \nu \nabla^2 \textbf{u}+\beta g \theta \hat{z}, \\
\frac{\partial \theta}{\partial t} + \textbf{u} \cdot \nabla \theta &=& \kappa \nabla^2 \theta,
\end{eqnarray}
for the velocity field $\textbf{u}$, the kinematic pressure field $p$, and the temperature field $\theta$ relative to some reference temperature. In the Boussinesq approximation it is assumed that the material properties of the fluid such as the thermal expansion coefficient $\beta $, the viscosity $\nu$, and the thermal diffusivity $\kappa$ do not depend on temperature. Here $g$ is the gravitational acceleration and $\boldsymbol \Omega$ is the rotation rate of the system around the center axis, pointing against gravity, $ \boldsymbol \Omega = \Omega {\bf \hat{z}}$. Note that $\rho(T_0)$ has been absorbed in the pressure term. In addition to the Oberbeck-Boussinesq equations we have the continuity equation ($\nabla \cdot \textbf{u} = 0$). A no-slip ($\textbf{u}=0$) velocity boundary condition is assumed at all the walls. Moreover, the hot bottom and the cold top plate are at a constant temperature, and no lateral heat flow is allowed at the sidewalls.

Within the Oberbeck-Boussinesq approximation and for a given cell geometry, the dynamics of the system is determined by three dimensionless control parameters, namely, the Rayleigh number 
\begin{equation}
Ra = \frac{\beta g \Delta L^3}{\kappa\nu},
\end{equation}
where $L$ is the height of the sample, $\Delta = T_b-T_t$ the difference between the imposed temperatures $T_b$ and $T_t$ at the bottom and the top of the sample, respectively, the Prandtl number
\begin{equation}
Pr = \frac{\nu} {\kappa},
\end{equation}
and the rotation rate which is indicated by the Rossby number
\begin{equation}
Ro =  \sqrt{\beta g \Delta /L}/(2\Omega).
\end{equation}
The Rossby number indicates the ratio between the buoyancy and Coriolis force. Note that the $Ro$ number is an inverse rotation rate. Alternative parameters to indicate the rotation rate of the system are the Taylor number
\begin{equation}
Ta = \left(\frac{2 \Omega L^2}{\nu}\right)^2,
\end{equation}
comparing Coriolis and viscous forces, or the Ekman number
\begin{equation}
Ek = \frac{\nu}{\Omega L^2}= \frac{2}{\sqrt{Ta}}.
\end{equation}
A convenient relationship between the different dimensionless rotation rates is $Ro=\sqrt{Ra/(Pr Ta)}$.

The cell geometry is described by its shape and an aspect ratio
\begin{equation}
	\Gamma=D/L,
\end{equation}
where $D$ is the cell diameter. The response of the system is given by the non-dimensional heat flux, i.e. the Nusselt number
\begin{equation}
    Nu= \frac{QL}{\lambda \Delta}~,
\label{eq:nusselt}
\end{equation}
where $Q$ is the heat-current density and $\lambda$ the thermal conductivity of the fluid, and a Reynolds number
\begin{equation}
Re = \frac{U L}{\nu}.
\end{equation}
There are various reasonable possibilities to choose a velocity $U$, e.g., the components or the magnitude of the velocity field at different positions, local or averaged amplitudes, etc, and several choices have been made by different authors. A summary of some work that has been done on rotating RB convection is given in section 2.8 of the book by Lappa. \cite{lap12}

\section{Parameter regimes covered}
\begin{figure*}
\centering
\subfigure{\includegraphics[width=0.44\textwidth]{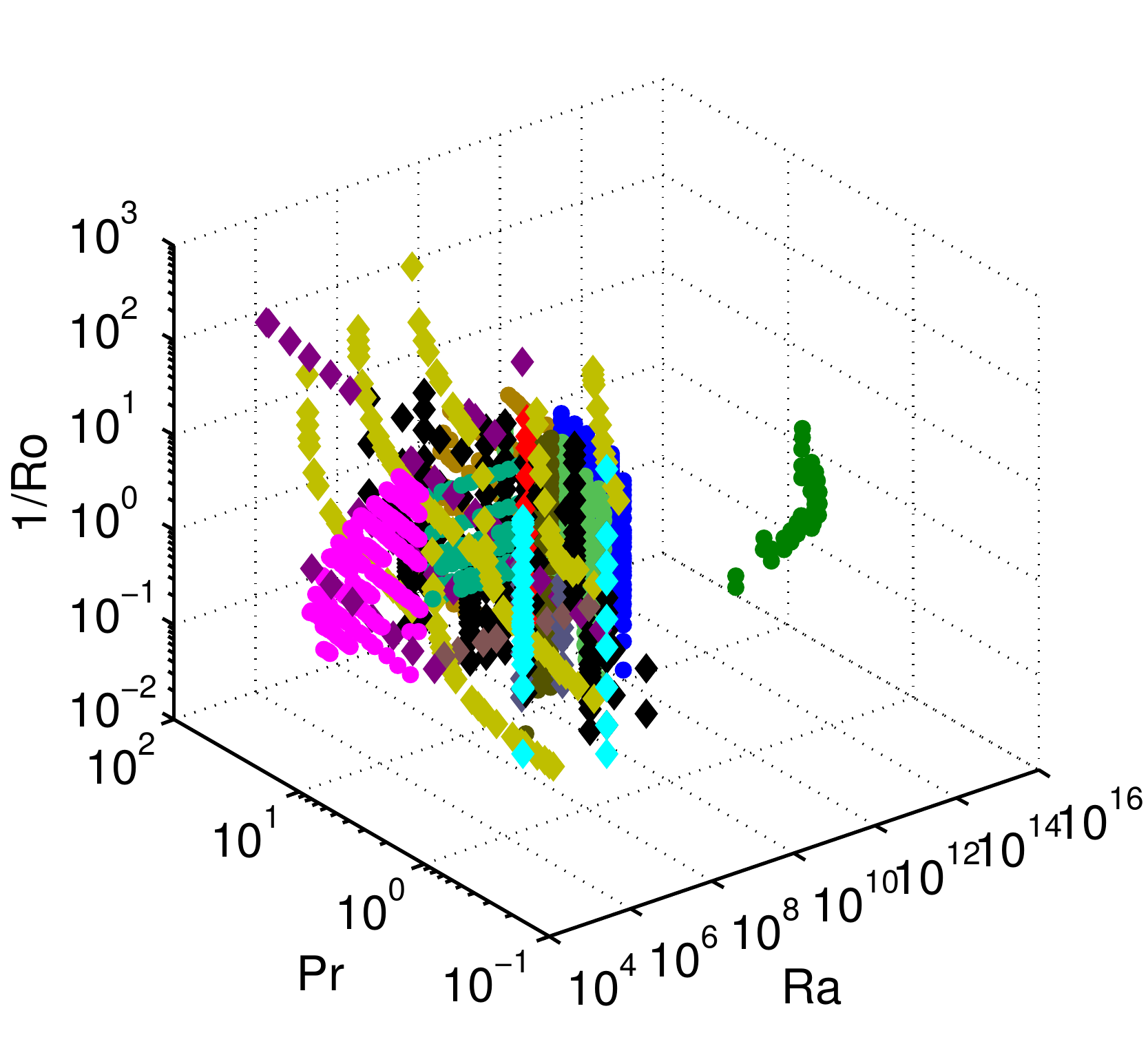}}
\subfigure{\includegraphics[width=0.44\textwidth]{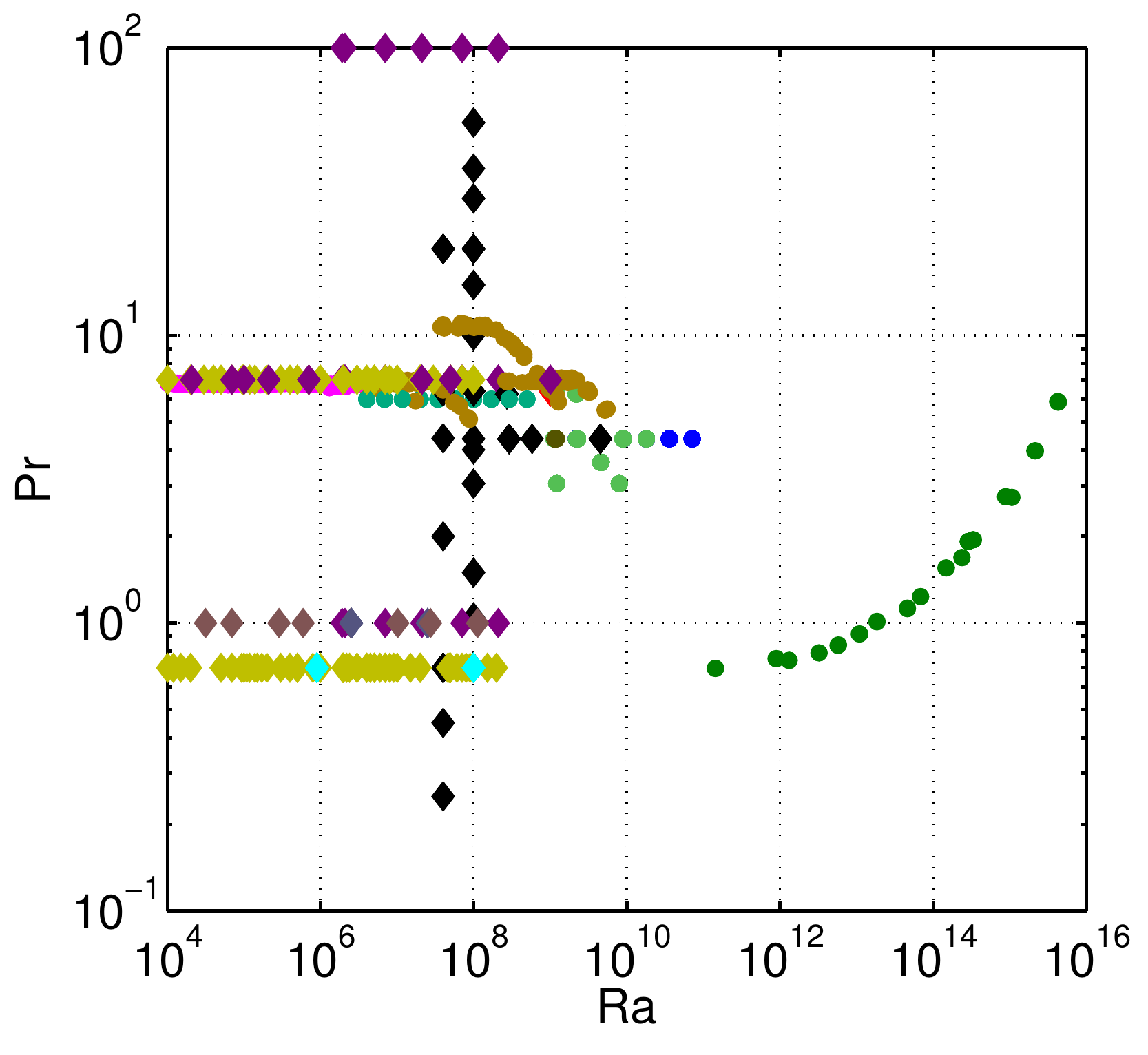}}
\subfigure{\includegraphics[width=0.44\textwidth]{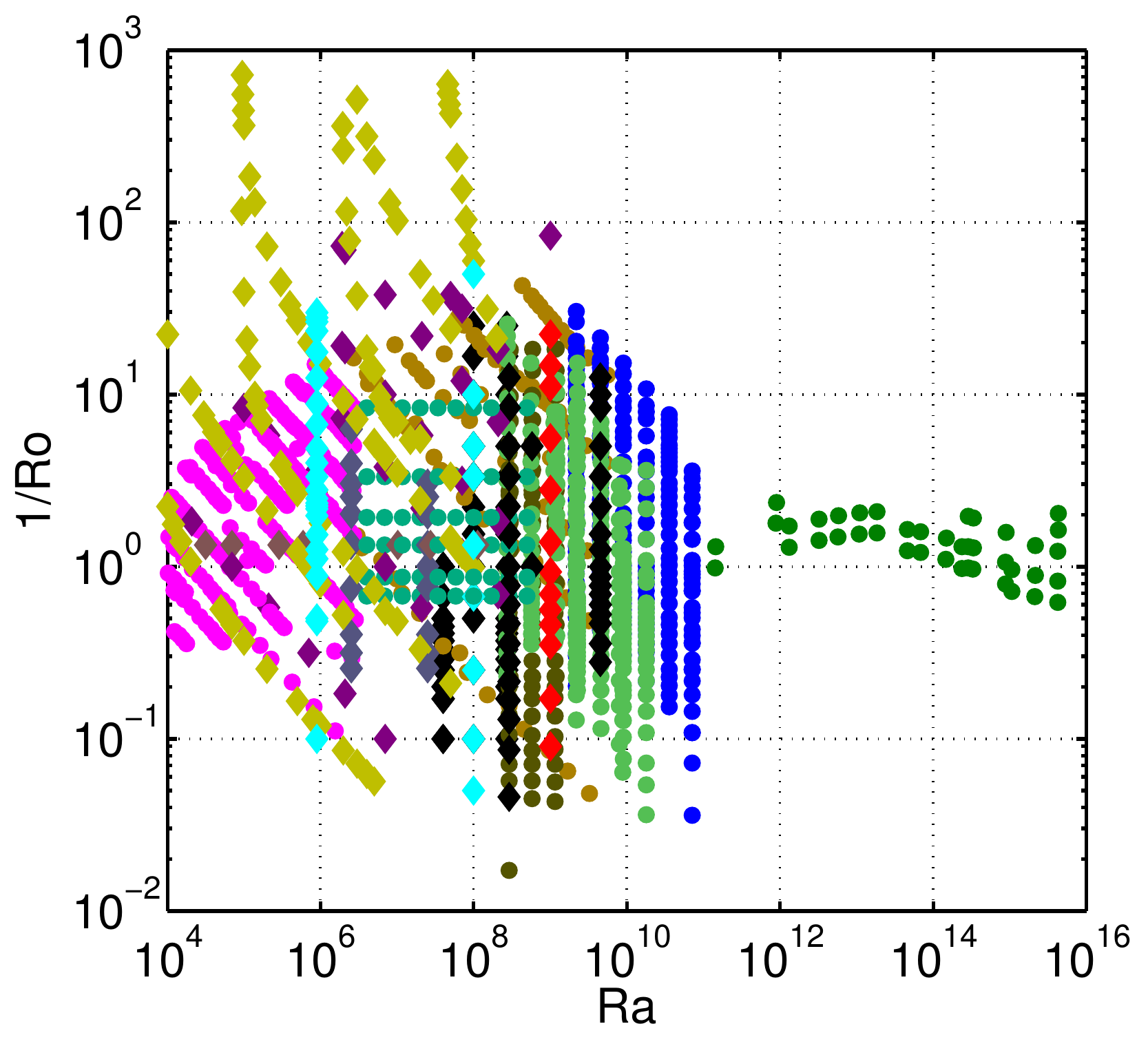}}
\subfigure{\includegraphics[width=0.44\textwidth]{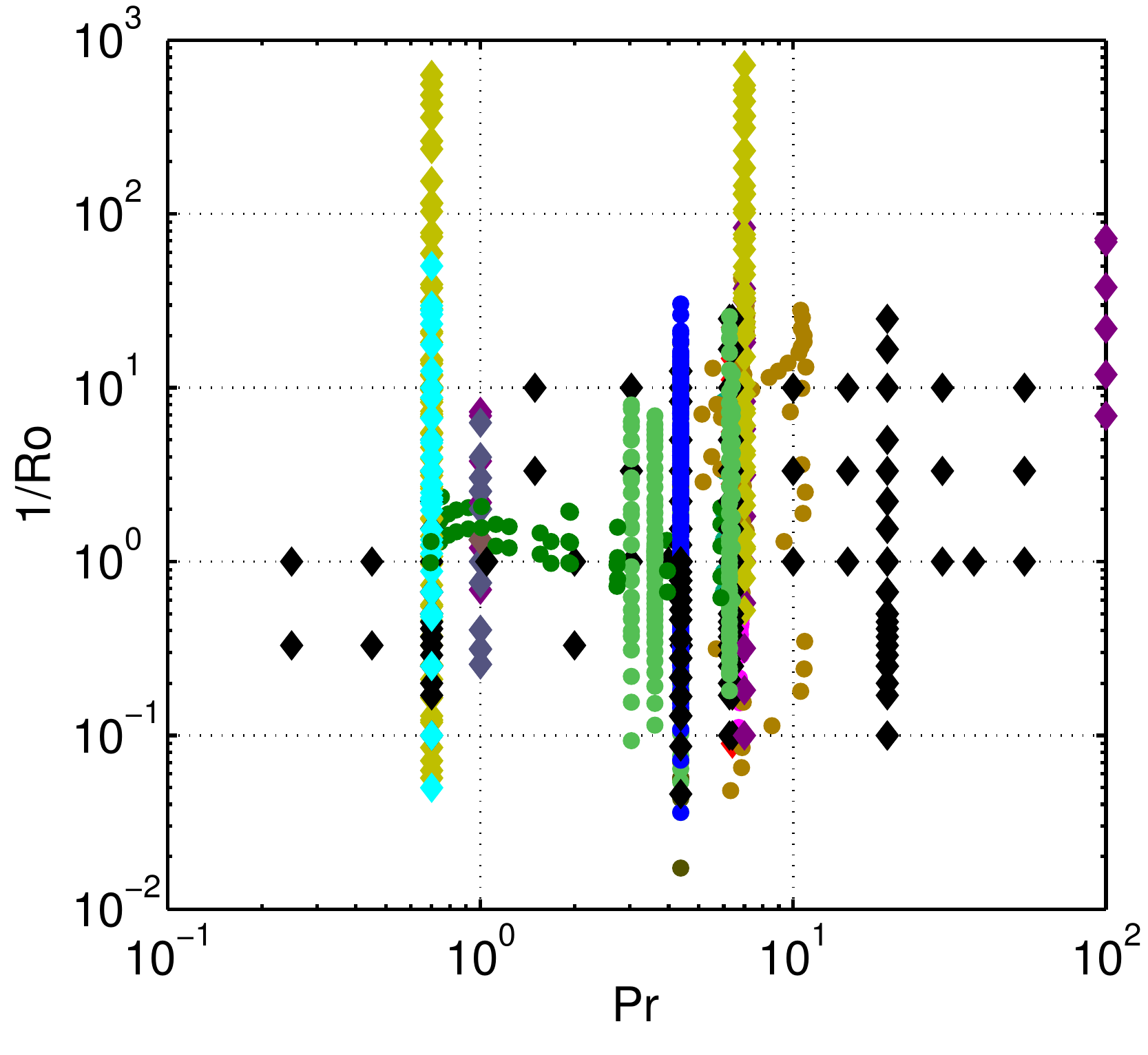}}
\subfigure{\includegraphics[width=0.77\textwidth]{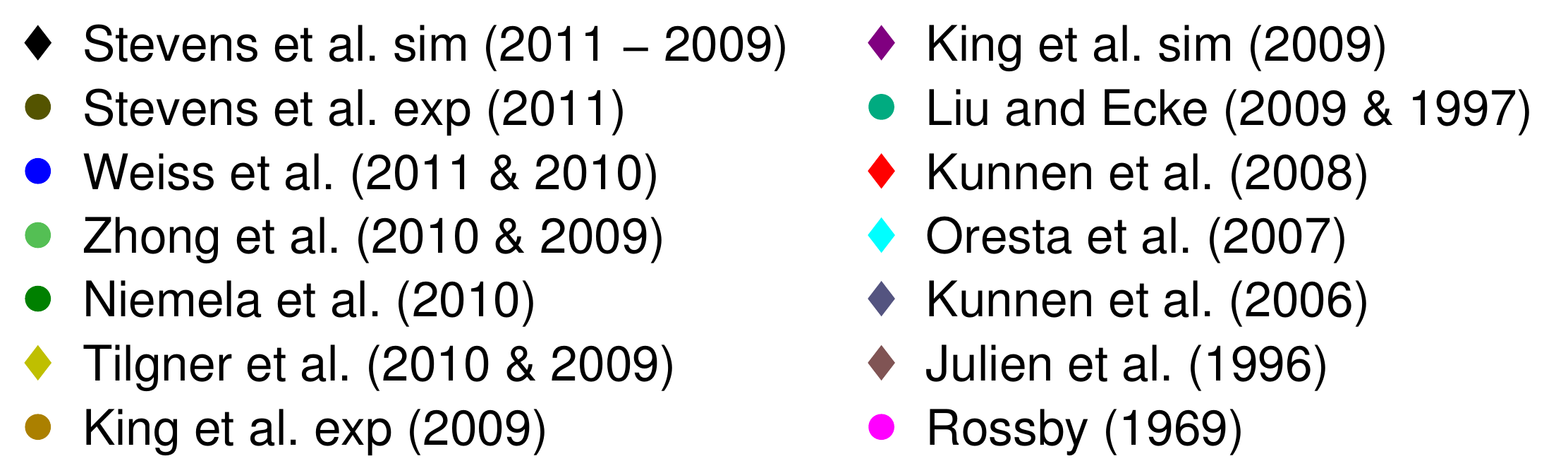}}
\caption{
The parameter diagram in Ra-Pr-Ro space for rotating RB convection; an updated version from Stevens {\it et al.}\ \cite{ste10b}. The data points indicate where $Nu$ has been measured or numerically calculated. The data are obtained in a cylindrical cell with aspect ratio $\Gamma=1$ with no slip boundary conditions, unless mentioned otherwise. The data from DNSs and experiments are indicated by diamonds and dots, respectively. The data sets are from:
 Stevens {\it et al.} (2011 - 2009) \cite{zho09b,ste09,wei10,ste10a,ste10b,ste11b} ($\Gamma=0.5$, $\Gamma=1.0$, $\Gamma=4/3$, $\Gamma=2.0$);
 Weiss $\&$ Ahlers (2011 $\&$ 2010) \cite{wei10,wei11,wei11b} ($\Gamma=0.5$) ;
 Zhong $\&$ Ahlers (2010 $\&$ 2009) \cite{zho09b,zho10c};
 Niemela {\it et al.} (2010) \cite{nie10} ($\Gamma=0.5$);
 Schmitz $\&$ Tilgner (2009 $\&$ 2010) \cite{sch09} (free slip and no-slip boundary conditions and horizontally periodic);
 King {\it et al.} (2009) (horizontally periodic and different aspect ratios) \cite{kin09};
 Liu $\&$ Ecke (2009 $\&$ 1997) \cite{liu09,liu97} (square with $\Gamma=0.78$);
 Kunnen {\it et al.} (2008) \cite{kun08b};
 Oresta {\it et al.} (2007) \cite{ore07} ($\Gamma=0.5$);
 Kunnen {\it et al.} (2006) \cite{kun06} ($\Gamma=2$, horizontally periodic);
 Julien {\it et al.} (1996) \cite{jul96} ($\Gamma=2$, horizontally periodic);
 Rossby (1969) (varying aspect ratio).
 Panel a shows a three-dimensional view on the parameter space (see also the movie in the supplementary material),
 b) projection on the Ra-Pr parameter space,
 c) projection on the Ra-Ro parameter space,
 d) projection on the Pr-Ro parameter space.
 }
\label{fig:figure1}
\end{figure*}

In figure \ref{fig:figure1} we present the explored $Ra-Pr-Ro$ parameter space for rotating RB convection \footnote{In figure 1 of Stevens et al. \cite{ste11b} also the $Ra-Ro-\Gamma$ parameter space for $Pr=4.38$ can be found.}. Here we emphasize that numerical simulations and experiments on rotating RB convection are complementary, because different aspects of the problem can be addressed. Namely, in accurate experimental measurements of the heat transfer a completely insulated system is needed. Therefore, one cannot visualize the flow while the heat transfer is measured. On the positive side, in experiments one can obtain very high $Ra$ numbers and long time averaging. In simulations, on the other hand, one can simultaneously measure the heat transfer while the complete flow field is available for analysis. But the $Ra$ number that can be obtained in simulations is lower than in experiments, due to the computational power that is needed to fully resolve the flow. Here we should also mention that the highest $Ra$ number reached in rotating RB experiments is almost $Ra=10^{16}$, whereas in DNSs of rotating RB convection the highest $Ra$ number is $Ra=4.52\times10^9$. However, the flexibility of simulations allows to study more $Pr$ numbers, i.e. covering a range of $0.2 \leq Pr \leq 100$, whereas present experiments are almost exclusively for $3.05 \leq Pr \leq 6.4$, i.e. the $Pr$ number regime accessible with water. In addition, we note that the $1/Ro$ number regime that can be covered in experiments can, depending on $Ra$ and $Pr$, be somewhat limited. Very low $1/Ro$ values, corresponding to very weak rotation, are difficult due to the accuracy limitations at very small rotation rates. The lowest rotation rates achieved in the recent Eindhoven \cite{kun08e,kun11,ste11b} and Santa Barbara \cite{zho10c,wei11} experiments are about 0.01 rad/s (one rotation every 10.5 minutes). For the Eindhoven experiments this is already rather close to the accuracy with which we can set the rotation rate. The highest $1/Ro$. The highest $1/Ro$ that can be obtained in a setup is either determined by the requirement that the Froude number $Fr=\Omega^2 (L/2) /g$ \cite{har00b}, which indicates the importance of centrifugal effects, is not too high (usually $Fr<0.05$ is considered to be a safe threshold \cite{zho09b}) or by the maximum rotation rate that can be achieved. 

\section{Nusselt number measurements}
Early linear stability analysis, see e.g. Chandrasekhar \cite{cha81}, revealed that rotation has a stabilizing effect due to which the onset of heat transfer is delayed. This can be understood from the thermal wind balance, which implies that convective heat transport parallel to the rotation axis is suppressed. Experimental and numerical investigations concerning the onset of convective heat transfer and the pattern formation in cylindrical cells just above the onset under the influence of rotation have been performed by many authors, see e.g.\ \cite{nak55,luc83,pfo84,pfo87,zho91,zho93,hu95,baj98,tag08,bor10,lop06,lop09,rub10,sch05b,sch10b}.

\begin{figure}[!t]
  \centering
  \includegraphics[width=0.49\textwidth]{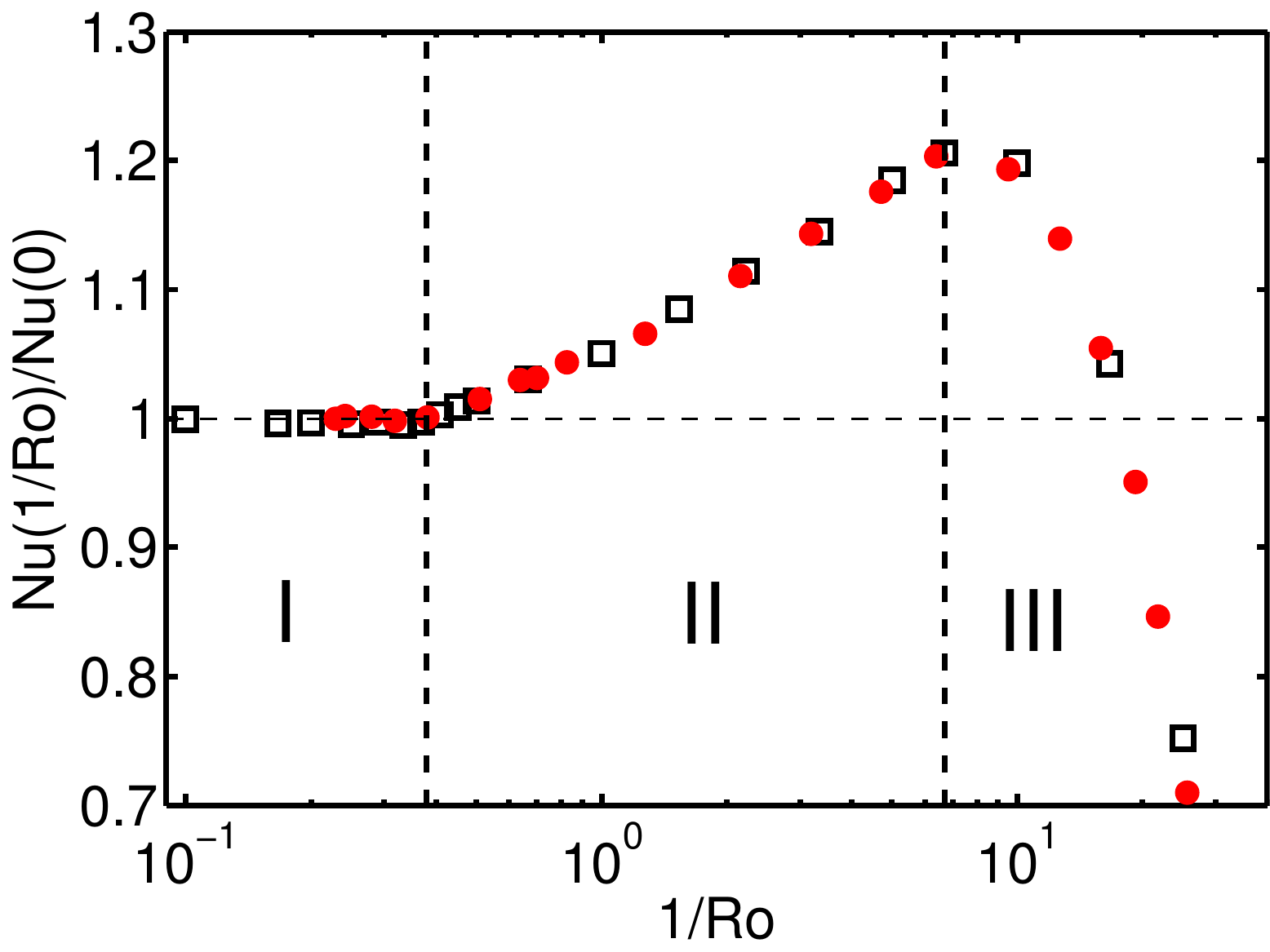}
  \caption{The scaled heat transfer $Nu(1/Ro)/Nu (0)$ as function of $1/Ro$ on a logarithmic scale. a) Experimental and numerical data for $Ra=2.73 \times 10^8$ and $Pr=6.26$ in a $\Gamma=1$ sample are indicated by red dots and open squares, respectively. Data taken from Stevens {\it et al.}\ \cite{ste09}.}
  \label{fig:figure2}
\end{figure}

\begin{figure}[!t]
\centering
\includegraphics[width=0.49\textwidth]{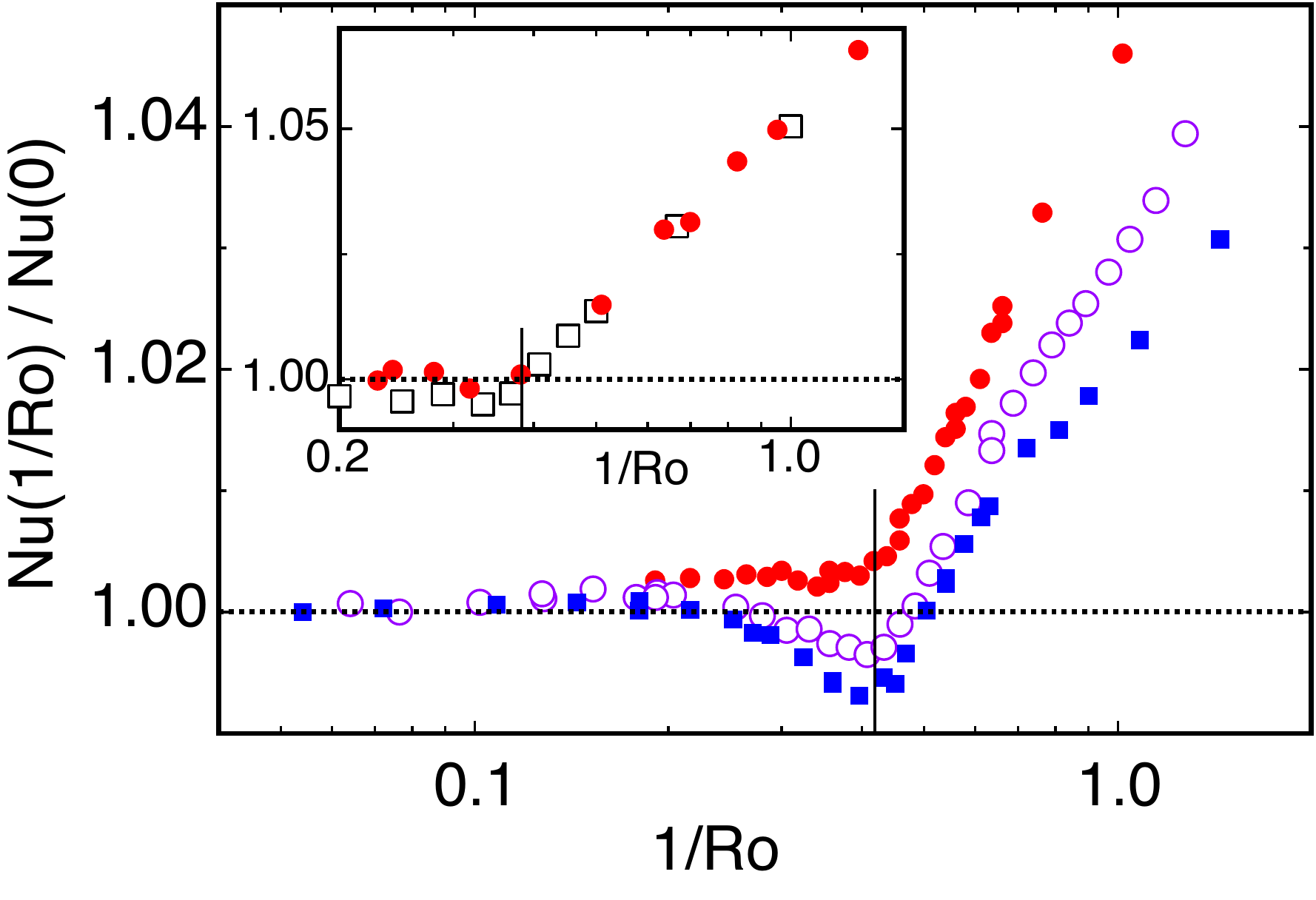}
\caption{
The Nusselt number $Nu(1/Ro)$, normalized by $Nu(0)$ without rotation, as a function of the inverse Rossby number $1/Ro$. Data are from experiments of Zhong and Ahlers \cite{zho10c}. Main figure: $\Gamma=1$, $Pr = 4.38$, $Ra = 2.25\times 10^9$ (solid circles), $8.97\times 10^9$ (open circles), and $1.79\times 10^{10}$ (solid squares). Inset: zoom in of figure \ref{fig:figure2}. Figure taken from Weiss {\it et al.}\ \cite{wei10}.
} 
\label{fig:figure3}
\end{figure}

Since the experiments by Rossby in 1969 \cite{ros69}, it is known that rotation can also enhance the heat transport. Rossby found that, when water is used as the convective fluid, the heat transport first increases when the rotation rate is increased. He measured an increase of about $10 \%$. This increase is counterintuitive as the stability analysis of Chandrasekhar \cite{cha81} has shown that rotation delays the onset to convection and from this one would expect that the heat transport decreases. The mechanism responsible for this heat transport enhancement is Ekman pumping \cite{ros69,jul96b,vor02,kun08b,kin09,zho09b}, i.e. due to the rotation, rising or falling plumes of hot or cold fluid are stretched into vertically-aligned-vortices that suck fluid out of the thermal boundary layers adjacent to the bottom and top plates. This process contributes to the vertical heat flux. For stronger rotation Rossby found, as expected, a strong heat transport reduction, due to the suppression of the vertical velocity fluctuations by the rotation. This means that a typical measurement of the heat transport enhancement with respect to the non-rotating case as function of the rotation rate looks like the one shown in figure \ref{fig:figure2}. After Rossby many experiments, e.g. \cite{zho93,jul96,liu97,vor02,kun08b,kun08e,kin09,liu09,zho09b,ste09,zho10c,kun10,kun10b,kun11,wei11,wei11b,kin12,pha11}, have confirmed this general picture. A new unexplained feature observed in high precision heat transport measurements is the small decrease in the heat transport just before the strong heat transport enhancement sets in, see figure \ref{fig:figure3}. The experiments reveal that this effect becomes stronger for higher $Ra$. Because only for $Ra\gtrsim 10^{10}$ the effect is more than $1\%$ it is currently not possible to study this in simulations, as at these high $Ra$ numbers the spatial and time resolution would be insufficient to capture such a small effect.

In early numerical simulations of rotating RB convection a horizontally unbounded domain was simulated. These simulations are relevant to understand large convective systems that are influenced by rotation, such as the atmosphere and solar convection. In 1991 Raasch $\&$ Etling \cite{raa91} used a large-eddy simulation to simulate the atmospheric boundary layer. Julien and coworkers \cite{jul96,jul96b,jul99,leg01}, Husain {\it et al.} \cite{hus06}, Kunnen {\it et al.}\ \cite{kun06,kun09} and King {\it et al.}\ \cite{kin09,kin12} used DNSs at several $Ra$ and $Ro$ to study the heat transport and the resulting flow patterns under the influence of rotation. Sprague {\it et al.}\ \cite{spr06} and Julien et al.\ \cite{jul12} numerically solved an asymptotically reduced set of equations valid in the limit of strong rotation. In this way Julien et al.\ \cite{jul12} identified as function of increasing $Ra$ four distinct regimes, i.e.\ a disordered cellular regime, a regime of weakly interacting convective Taylor columns, a disordered plume regime characterized by reduced heat transport efficiency, and finally geostrophic turbulence. Finally, Schmitz $\&$ Tilgner \cite{sch09,sch10} performed horizontally periodic simulations with no-slip and stress-free boundary conditions at the horizontal plates. They show that heat transport enhancement under the influence of rotation is only found when a no-slip boundary condition at the horizontal plates is used. In recent years the focus of numerical simulations has shifted from horizontally periodic simulations to rotating RB convection in a cylindrical sample in order to make a direct comparison with experiments possible \cite{ore07,kun08b,kun10,kun10b,kun11,zho09b,wei10,ste09,ste10a,ste10b,ste11b,ste12b,hor11}. As all information on the flow field is available in these simulations these studies focus on determining the influence of rotation on the heat transport and the corresponding changes in the flow structure.

\begin{figure}[!t]
\centering
\subfigure{\includegraphics[width=0.49\textwidth]{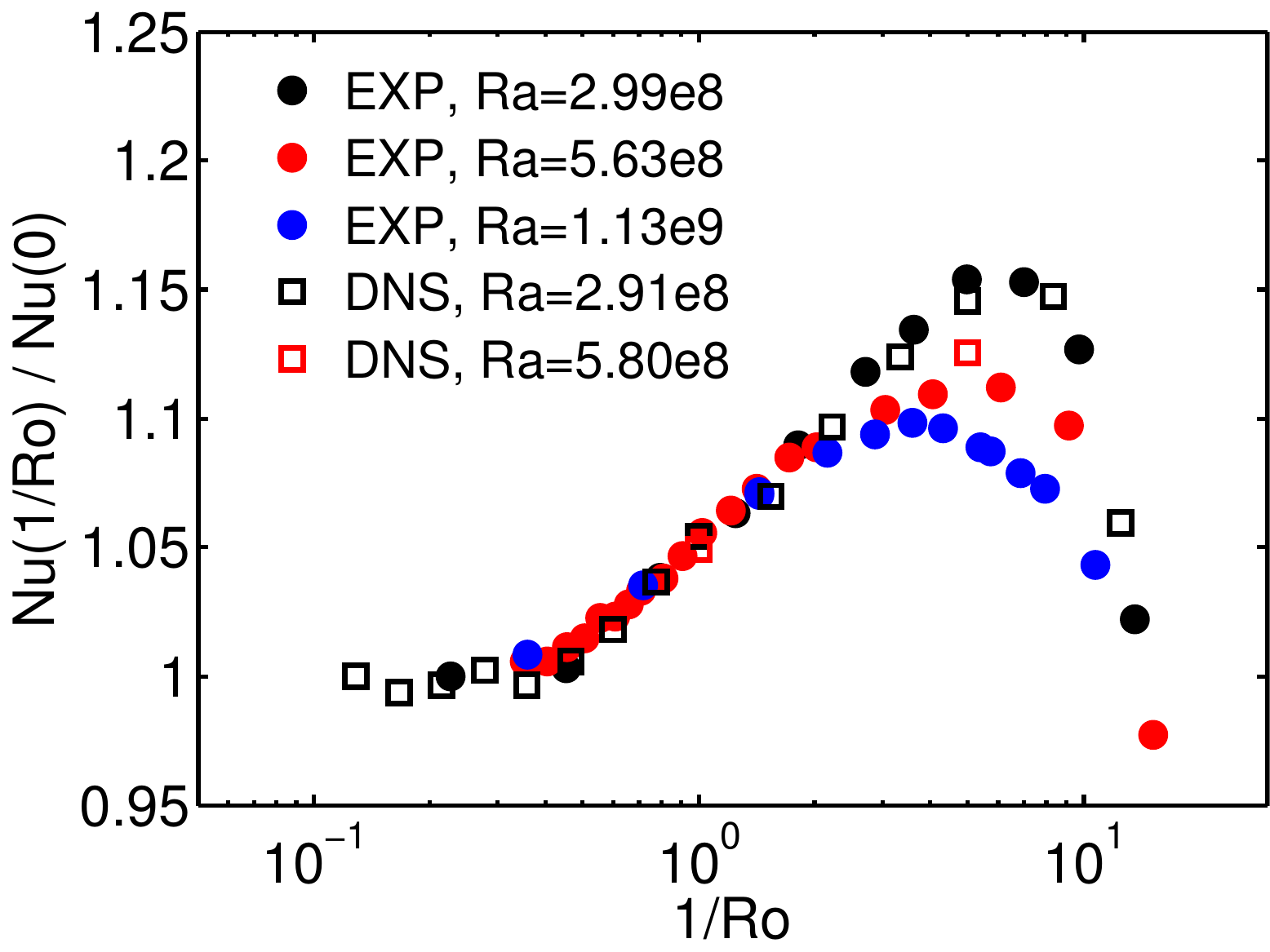}}
\caption{The ratio $Nu(1/Ro)/Nu(0)$ as function of $1/Ro$ for different $Ra$ in a $\Gamma=1$ sample. The experimental results for $Ra=2.99\times10^8$ (Stevens {\it et al.}\ \cite{ste11b}), $Ra=5.63\times10^8$ (Zhong $\&$ Ahlers \cite{zho10c}), and $Ra=1.13\times10^9$ (Zhong $\&$ Ahlers \cite{zho10c}) are indicated in black, red, and blue solid circles, respectively. The DNS results from Stevens {\it et al.}\ \cite{ste11b} for $Ra=2.91 \times10^8$, and $Ra=5.80\times10^8$ are indicated by black and red open squares, respectively. All presented data in this figure are for $Pr=4.38$. Figure taken from Stevens {\it et al.}\ \cite{ste11b}.
}
\label{fig:figure4}
\end{figure}

\begin{figure}[!t]
\centering
\includegraphics[width=0.49\textwidth]{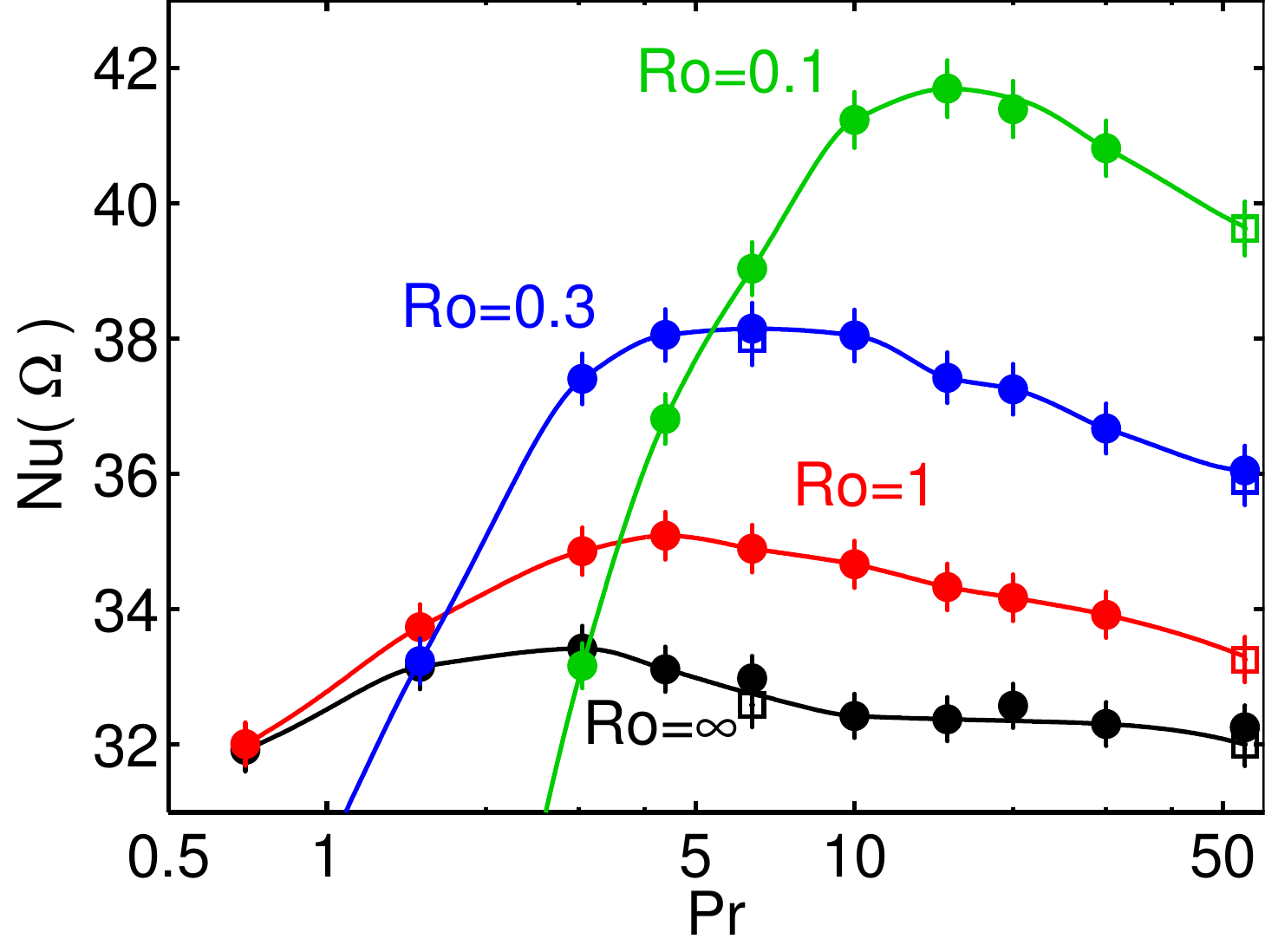}
\caption{The heat transfer as function of $Pr$ on a logarithmic scale for $Ra = 1\times 10^8$. Black, red, blue, and green indicate the results for $Ro=\infty$, $Ro=1.0$, $Ro=0.3$, and $Ro=0.1$, respectively. The data obtained on the $385\times 193 \times 385$ (open squares) and the $257\times 129 \times 257$ grid (solid circles) are in very good agreement. Figure taken from Stevens {\it et al.}\ \cite{ste10a} }
\label{fig:figure5}
\end{figure}

A direct comparison between experiments and simulations was used by Zhong {\it et al.} \cite{zho09b} and Stevens {\it et al.} \cite{ste10a} to study the influence of $Ra$ and $Pr$ on the effect of Ekman pumping. A strong heat transport enhancement with respect to the non-rotating case was observed for $Pr\approx 6$  \cite{zho09b}. The maximum enhancement decreases with increasing $Ra$ and decreasing $Pr$, see figure \ref{fig:figure4} and \ref{fig:figure5}. Later Stevens {\it et al.} \cite{ste10a} found that at a fixed $Ro$ number the effect of Ekman pumping, and thus the observed heat transport enhancement with respect to the non-rotating case, is highest at an {\it intermediate} Prandtl number, see figure \ref{fig:figure5}. At lower $Pr$ the effect of Ekman pumping is reduced as more hot fluid that enters the vortices at the base spreads out in the middle of the sample due to the large thermal diffusivity of the fluid. At higher $Pr$ the thermal boundary layer becomes thinner with respect to the kinematic boundary layer, where the base of the vortices is formed, and hence the temperature of the fluid that enters the vortices becomes lower. In this framework the decrease of the heat transport enhancement at higher $Ra$ can be explained by the increase of the turbulent viscosity with increasing $Ra$. These observations also explain why no heat transport enhancement due to rotation is observed in the very high $Ra$ number experiments of Niemela {\it et al.} \cite{nie10}. In recent simulations of Horn {\it et al.} \cite{hor11} the effect of non-Oberbeck-Boussinesq (NOB) corrections was studied in rotating RB convection and they showed that for water NOB corrections leads to a Nusselt number that is a few percent higher than for the Oberbeck-Boussinesq case. 

\section{Different turbulent states}

\begin{figure*}[!t]
\centering
\subfigure{\includegraphics[width=0.49\textwidth]{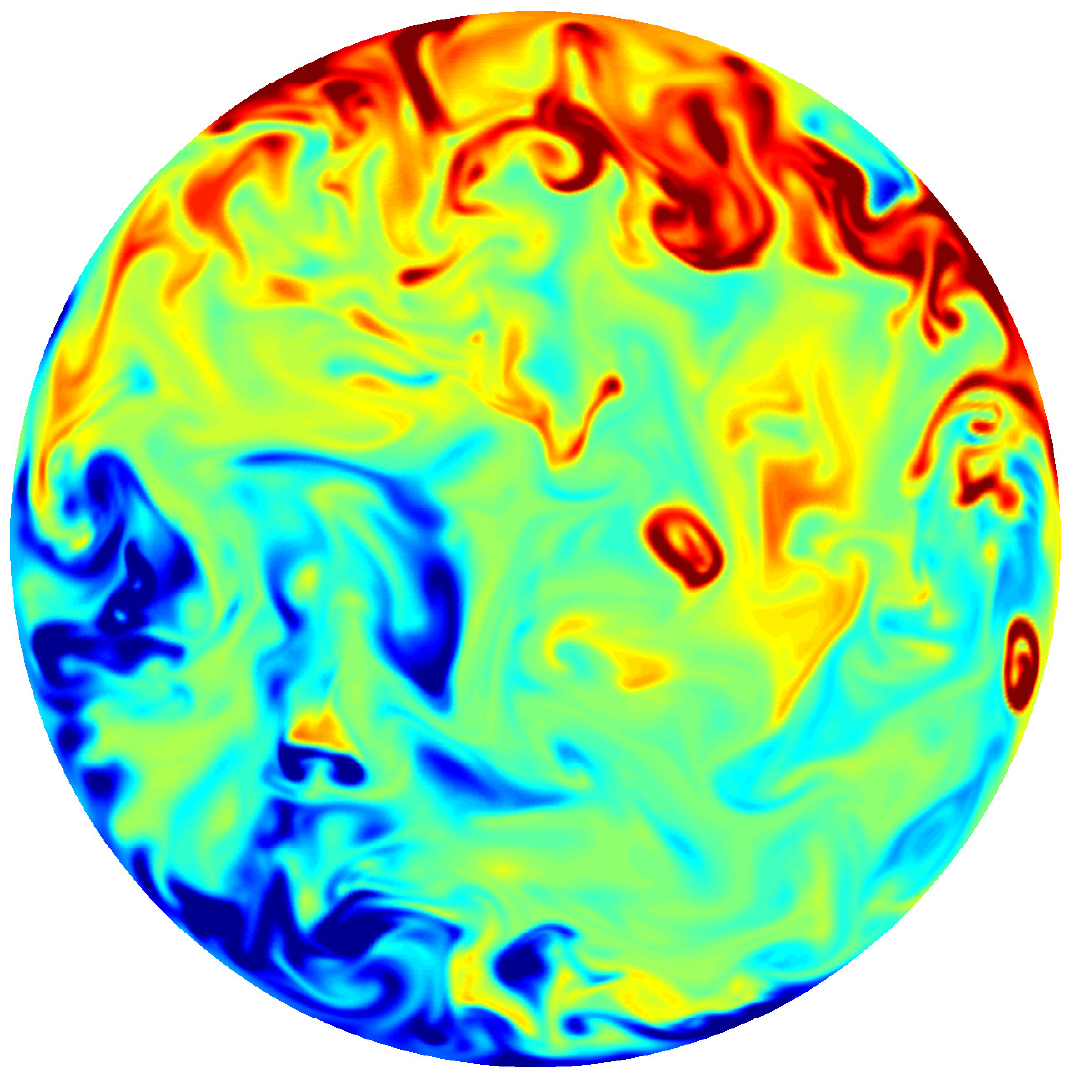}}
\subfigure{\includegraphics[width =0.49\textwidth]{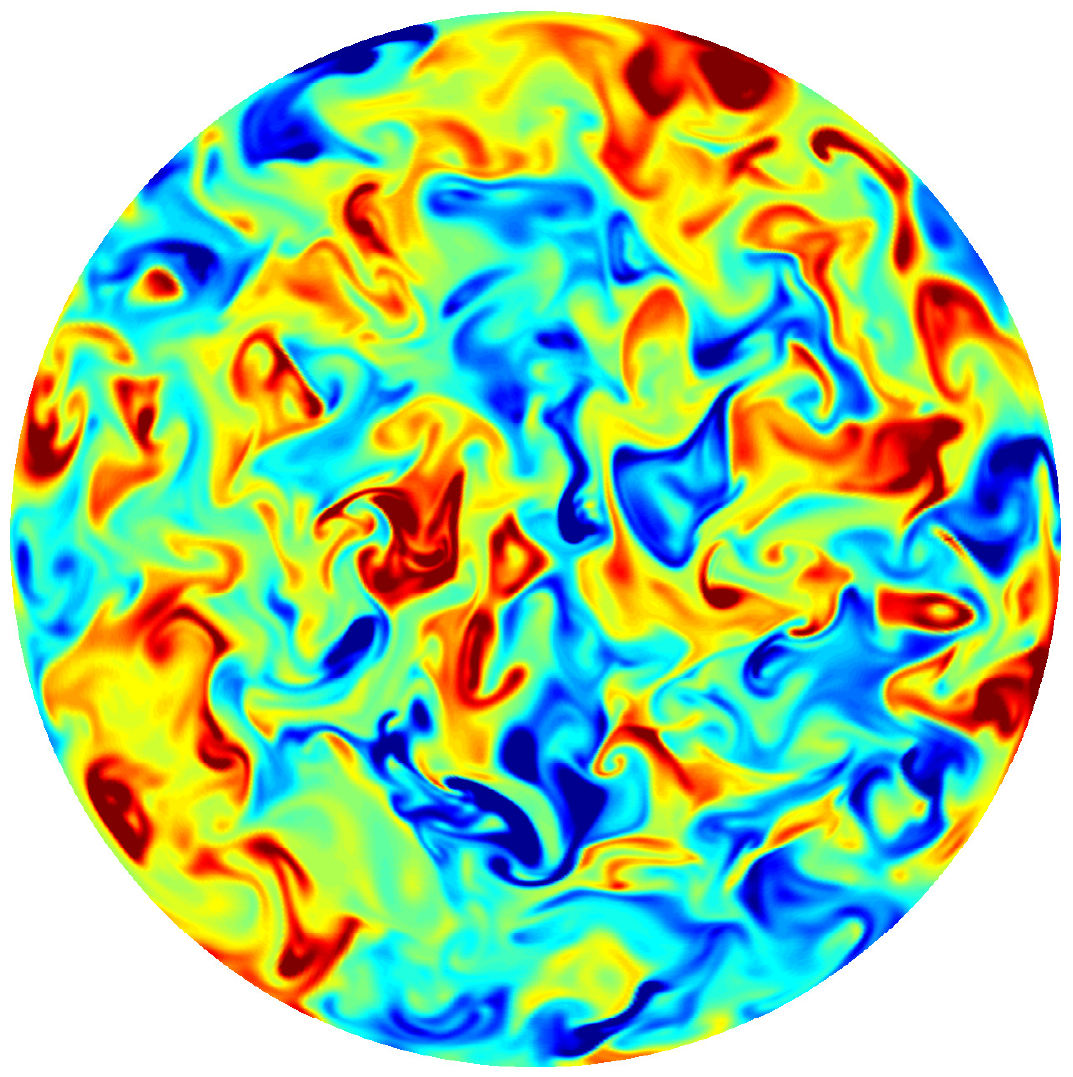}}
\caption{Visualization, based on a DNS at $Ra=2.91 \times 10^8$ with $Pr=4.38$ in a $\Gamma=2$ sample, of the temperature field in the horizontal mid-plane of a cylindrical convection cell. The red and blue areas indicate warm and cold fluid, respectively. The left image is at a slow rotation rate ($1/Ro=0.046$) below the transition where the warm upflow (red) and cold downflow (blue) reveal the existence of a single convection roll superimposed upon turbulent fluctuations. The right image is at a somewhat larger rotation rate ($1/Ro=0.6$) and above the transition, where vertically-aligned vortices cause disorder on a smaller length scale.}
\label{fig:figure6}
\end{figure*}

When the heat transport enhancement as function of the rotation rate is considered, a division in three regimes is possible \cite{bou90,kun10,kun11}. Here we will call these regimes: regime I (weak rotation), regime II (moderate rotation), and regime III (strong rotation), see figure \ref{fig:figure2}. It is well known that without rotation a large scale circulation (LSC) is the dominant flow structure in RB convection, see e.g.\ Ref.\ \cite{ahl09} and figure \ref{fig:figure6}a. This has motivated Hart, Kittelman $\&$ Ohlsen \cite{har02}, Kunnen {\it et al.}\ \cite{kun08b}, and Ahlers and coworkers \cite{bro06,zho10c,wei11b} to study the influence of weak rotation on the LSC. In these investigations it was found that the LSC has a weak precession under the influence of rotation. In addition, it was shown by Zhong $\&$ Ahlers \cite{zho10c} that, although the Nusselt number is nearly unchanged in regime I (see figure \ref{fig:figure2}), there are various properties of the LSC that are changing significantly when the rotation rate is increased in regime I. Here we mention the increase in the temperature amplitude of the LSC, the LSC precession (also observed by Hart {\it et al.}\ \cite{har02} and Kunnen {\it et al.}\ \cite{kun08b}), the decrease of the temperature gradient along the sidewall, and the increased frequency of cessations. To our knowledge these observations are still not fully understood, but recently Assaf {\it et al.} \cite{ass12} developed and used a low-dimensional stochastic model, in the spirit of the one developed by Brown and Ahlers \cite{bro07}, that well describes the cessation frequencies measured in the experiments by Zhong and Ahlers \cite{zho10c}.

It was shown by Stevens {\it et al.}\ \cite{ste09} that the heat transport enhancement at the start of regime II sets in as a sharp transition, see inset figure \ref{fig:figure2}, due to the transition from a flow state dominated by a LSC at no or weak rotation to a regime dominated by vertically-aligned vortices after the transition. They demonstrate that in experiments the transition between the two states is indicated by changes in the time averaged LSC amplitudes, i.e. the average amplitude of the cosine fit to the azimuthal temperature profile measured with probes in the sidewall, and the temperature gradient at the sidewall. Later experiments and simulations by Weiss $\&$ Ahlers \cite{wei11b} and Kunnen {\it et al.}\ \cite{kun11} confirmed that also other statistical quantities of the sidewall measurements change at the onset of heat transport enhancement. In addition, the simulations by Stevens {\it et al.}\ \cite{ste09} and Weiss {\it et al.}\ \cite{wei10} show a strong increase in the number of vortices at the thermal boundary layer height, when the heat transport enhancement sets in. It was also revealed that at this point the character of the kinematic boundary layer changes from a Prandtl-Blasius boundary layer at no or weak rotation to an Ekman boundary layer after the onset, which is revealed by the ${Ro}^{1/2}$ scaling of the kinematic boundary layer thickness after the onset, see figure \ref{fig:figure7}. Furthermore, a strong increase in the vertical velocity fluctuations at the edge of the thermal boundary layer, due to the effect of Ekman pumping, was observed, while the volume  averaged value decreased due to the destruction of the LSC.

Here we also mention that there have been analytic modeling efforts by Julien {{\it et al.}\ \cite{jul07}, Portegies {\it et al.}\ \cite{por08} and Grooms {\it et al.}\ \cite{gro10} to understand the heat transport in regime III. In these models only the heat transfer in the vertically-aligned vortices is considered as most heat transport in this regime takes place inside these vortices. The trends shown by these models are in good agreement with simulation results. Furthermore, it was shown by Kunnen {\it et al.}\ \cite{kun11} that under the influence of rotation a secondary flow is created in regime II and III. This flow is driven by the Ekman boundary layers near the plates and generates a recirculation in the Stewartson boundary layer on the sidewall with upward (downward) transport of hot (cold) fluid close to the sidewall in the bottom (top) part of the cell. Hence this secondary flow leads to the generation of a vertical temperature gradient along the sidewall, which is measured in experiments \cite{kun11,zho09b,wei11b,zho10c} and simulations \cite{kun11}. We note that this secondary circulation is significantly different from that suggested by Hart {\it et al.}\ \cite{har99}, based on a linear mean temperature gradient, or that of Homsy $\&$ Hudson \cite{hom69}. Under the influence of strong rotation, i.e. for small values of $Ro$, a destabilizing temperature gradient is also formed in the bulk. This temperature gradient is caused by the merger of the vertically-aligned plumes \cite[see, e.g.,][]{bou86,zho93,jul96,eck98,jul99,leg01,spr06}, i.e. the enhanced horizontal mixing of the temperature anomaly of the plumes results in a mean temperature gradient in the bulk. Recently, this effect was measured for different $1/Ro$ and $Ra$ by Liu $\&$ Ecke \cite{liu11} with local temperature measurements in the bulk, which show good agreement with earlier simulations \cite{zho09b} and experiments by Hart {\it et al.}\ \cite{har99}.

\begin{figure}[!t]
  \centering
  \subfigure{\includegraphics[width=0.49\textwidth]{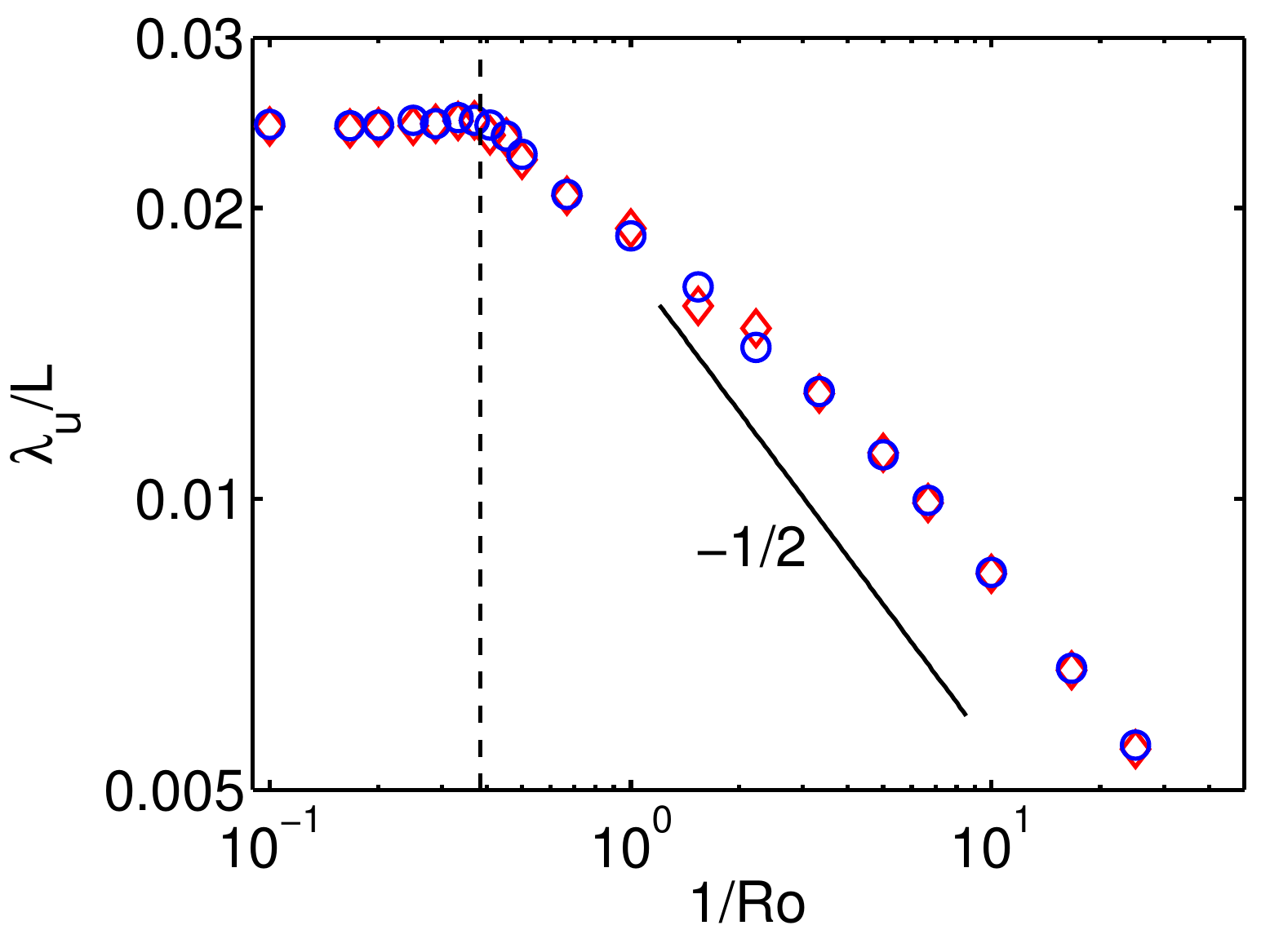}}
  \caption{The thickness of the kinematic top (diamonds) and bottom (circles) boundary layer for $Ra=2.73 \times 10^8$ and $Pr=6.26$ in a $\Gamma=1$ sample. The black line indicates the Ekman BL scaling of $(1/Ro)^{-1/2}$. Figure taken from Stevens {\it et al.}\ \cite{ste10b}.}
  \label{fig:figure7}
\end{figure}

\section{Sidewall temperature measurements}
In recent high precision heat transport measurements of rotating RB convection the samples are equipped with 24 thermistors embedded in the sidewall \cite{zho09b,zho10c,wei10,wei11b,kun11}. These thermistors are divided over 3 rings of 8 thermistors that are placed at $0.25 z/L$, $0.50 z/L$, and $0.75 z/L$. In non-rotating convection this arrangement of thermistors is used to determine the orientation of the LSC as the thermistors can detect the location of the upflow (downflow) by registering a relatively high (low) temperature. This method has been very successful in studying the statistics of the LSC in non-rotating RB convection, see the review of Ahlers {\it et al.}\ \cite{ahl09}. 

To determine the strength and orientation of the LSC a cosine function is fitted to the azimuthal temperature profile obtained from the sidewall measurements \cite{ahl09} as 
\begin{equation}\label{Eq cosine fit}
    \theta_i = \theta_k + \delta_k \cos(\phi_i-\phi_k).
\end{equation}
Here $\phi_i=2 i\pi/n_p$, where $n_p$ is the number of probes, indicates the azimuthal position of the probes, while $\theta_k$ gives the mean temperature, $\delta_k$ the temperature amplitude of the LSC, and $\phi_k$ the orientation. The index $k$ indicates whether the data are considered at $z/L=0.25$ (b), $z/L=0.50$ (m) or $z/L=0.75$ (t). In simulations similar measurements are obtained by Stevens {\it et al.}\ \cite{ste12b} and Kunnen {\it et al.}\ \cite{kun08b,kun11} from numerical probes placed close to the sidewall.

In rotating RB convection it is important to know when the LSC	 breaks down. Zhong {\it et al.}\ \cite{zho09b} showed that at the onset of heat transport enhancement the temperature amplitude of the LSC decreases, while an increase in the vertical temperature gradient at the sidewall is observed. Kunnen {\it et al.} \cite{kun08b} used simulation data to show that the energy in the first Fourier mode of the azimuthal temperature profile at the sidewall decreases strongly when the heat transport enhancement sets in. This idea was quantified by Stevens {\it et al.} \cite{ste10c} who proposed to determine the relative LSC strength based on the energy in the different Fourier modes of the measured or computed azimuthal temperature profile at or nearby the sidewall, as
\begin{equation}\label{Eq Relative Strength LSC}
    \bar{S}_k = \mathrm{max}\left( \left( \frac{\int_{t_b}^{t_e} E_1 dt}{\int_{t_b}^{t_e} E_{tot} dt} - \frac{1}{N}\right) / \left(1-\frac{1}{N}\right), 0 \right).
\end{equation}
The index $k$ has the same meaning as indicated below eq.\ (\ref{Eq cosine fit}). In eq.\ (\ref{Eq Relative Strength LSC}) $\int_{t_b}^{t_e} E_1 dt$ indicates energy in the first Fourier mode over the time interval $[t_b,t_e]$ of the simulation or experiment, and $\int_{t_b}^{t_e} E_{tot} dt$ the total energy in all Fourier modes over the same time interval. $N$ indicated the total number of Fourier modes that can be determined from the number of azimuthal probes that is available. From the definition of the relative LSC strength it follows that always $0\le \bar{S}_k \le 1$. Concerning the limiting values: $\bar{S}_k=1$ means the presence of a pure azimuthal cosine profile and $\bar{S}_k=0$ indicates that the magnitude of the cosine mode is equal to (or weaker than) the value expected from a random noise signal. We consider a value for $\bar{S}_k$ of about $0.5$ or higher as  an indicated that a cosine fit on average is a reasonable approximation of the data set and $\bar{S}_k$ well below $0.5$ that most energy is in the higher Fourier modes. It was shown by Kunnen {\it et al.}\ \cite{kun11}, Stevens {\it et al.}\ \cite{ste12b}, and Weiss {\it et al.}\ \cite{wei11b} that in a $\Gamma=1$ sample the relative LSC strength is close to $1$ before the onset of heat transport enhancement and strongly decreases to values close to zero at the moment that heat transport enhancement sets in, thus confirming that the LSC breaks down at this moment, see figure \ref{fig:figure8}a. 

\begin{figure*}[!t]
  \centering
  \subfigure{\includegraphics[width=0.48\textwidth]{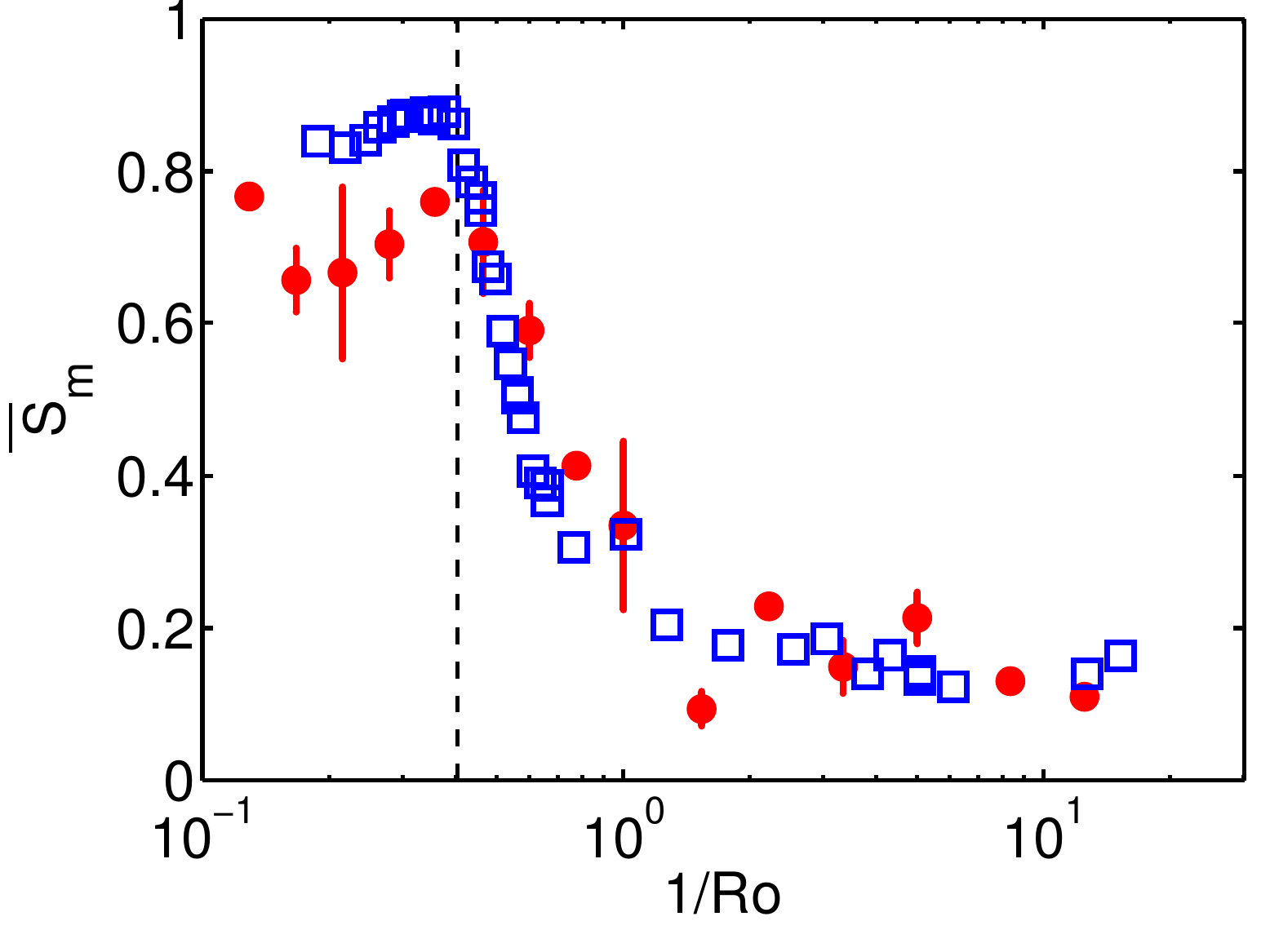}}  
  \subfigure{\includegraphics[width=0.48\textwidth]{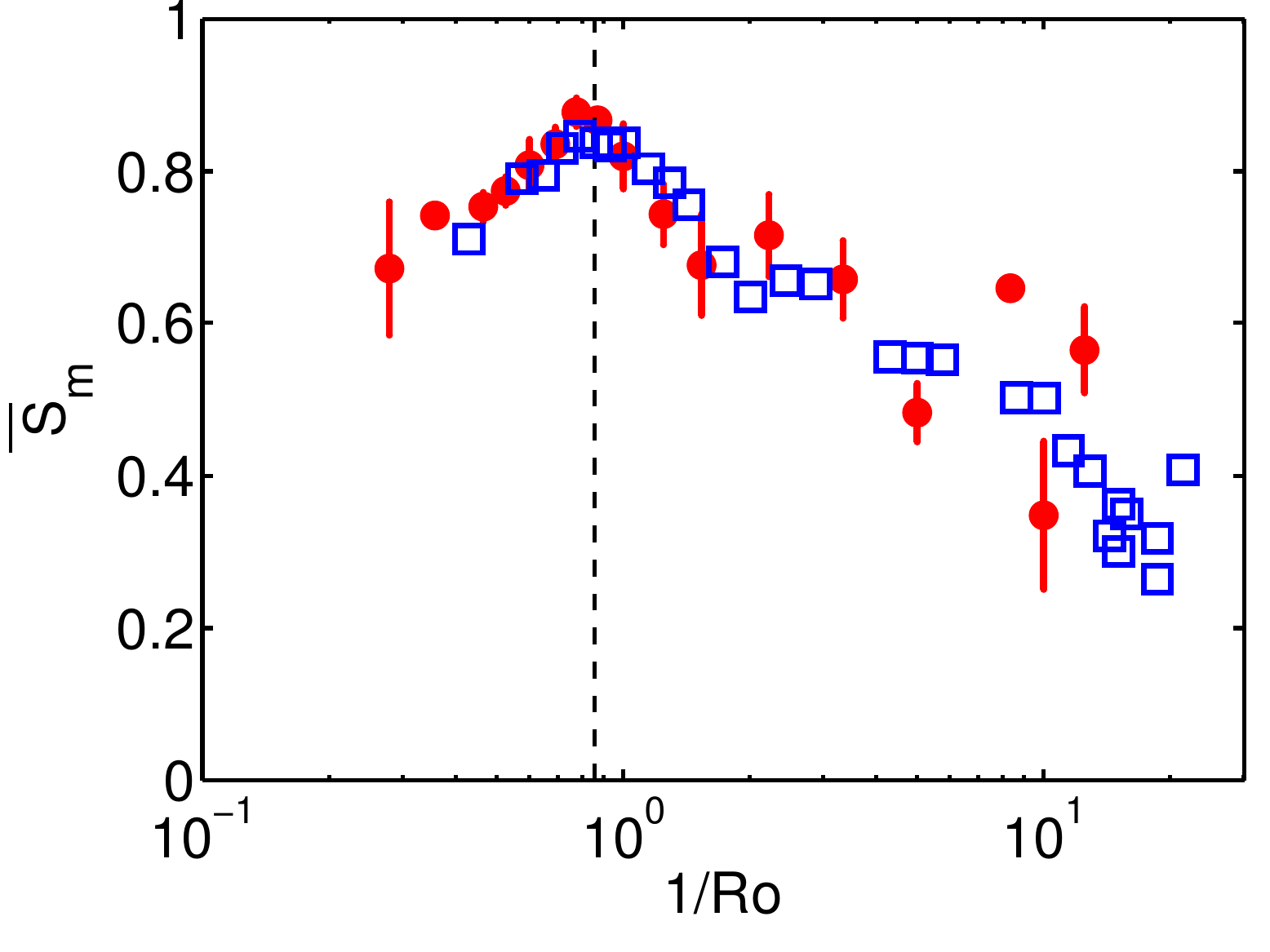}}
  \caption{The relative LSC strength at $0.50z/L$, i.e. $\bar{S}_m$, as function of $1/Ro$ based on experimental data (open squares) and simulations (filled circles). Panel (a) shows experimental data from Zhong $\&$ Ahlers \cite{zho10c} at $Ra=2.25 \times 10^9$ and simulation results from Stevens {\it et al.}\ \cite{ste12b} at $Ra=2.91\times10^8$ in a $\Gamma=1$ sample. Panel (b) shows experimental data from Weiss $\&$ Ahlers \cite{wei11b} and simulation results from Stevens at al.\ \cite{ste12b} for $Ra=4.52\times10^9$ in a $\Gamma=1/2$  sample. All presented data are for $Pr=4.38$. The vertical dashed line indicates the position where heat transport enhancement starts to set in, i.e. at $1/Ro_c\approx 0.86$ ($1/Ro_c\approx 0.40$) for $\Gamma=1/2$ ($\Gamma=1$) according to Weiss {\it et al.}\ \cite{wei10}. Figure taken from Stevens {\it et al.}\ \cite{ste12b}}
  \label{fig:figure8}
\end{figure*}

\section{Determination of vortex statistics}
In regime II and III, see figure \ref{fig:figure2}, the flow is dominated by vertically-aligned vortices. The experiments of Boubnov $\&$ Golitsyn \cite{bou90}, Zhong, Ecke $\&$ Steinberg \cite{zho93}, and Sakai \cite{sak97} first showed with flow visualization experiments that there is a typical spatial ordering of vertically-aligned vortices under the influence of rotation. In general the vertically-aligned vortices prefer to arrange themselves in a checkerboard pattern. This is nicely visualized in figure 1 of Ref.\ \cite{gro10}. A three-dimensional visualization of the vortices is made by Kunnen {\it et al.}\ \cite{kun10b} and Stevens {\it et al.} \cite{ste12b}, also based on numerical simulations. The first experimental measurements of velocity and temperature fields in rotating RB samples have been reported at least 20 years ago, and some of the pioneering work has been reported by Boubnov $\&$ Golitsyn \cite{bou90} and Fernando {\it et al.}\ \cite{fer91}. In both experiments the flow was analyzed using particle-streak photography. Later Vorobieff $\&$ Ecke \cite{vor98,vor02} used digital cameras to perform particle image velocimetry (PIV) measurements to study the flow with higher spatial and temporal resolutions. In addition, they added thermochromic liquid crystal (TLC) microcapsules to visualize the vortices in the temperature field. In later experiments by Kunnen {\it et al.}\ \cite{kun08b,kun08e,kun10} flow measurements were extended by the application of stereoscopic PIV (SPIV): In this technique two cameras are used to obtain a stereoscopic view, which allows for a reconstruction of the out-of-plane velocity. This made new experimental data available, for example on the correlation between the vertical velocity and vorticity. Recently, also King {\it et al.}\ \cite{kin12b} used flow visualization experiments to confirm the formation of vortices in the rotating regime. 

Although the presence of vortices in many cases is clear from snapshots of the flow, a well-defined vortex detection criteria are necessary to study the vortex statistics. Common methods are based on the velocity gradient tensor ${\bf \nabla u} = \partial_i u_j (i,j \in {1,2,3})$ \cite{hun88,hal05}. This tensor can be split into a symmetric and antisymmetric part 
\begin{equation}\label{}
    {\bf \nabla u} = \frac{1}{2} [ {\bf \nabla u} + ({\bf \nabla u})^T] + \frac{1}{2} [{\bf \nabla u} - ({\bf \nabla u})^T]= S + \Omega.
\end{equation}
The $Q_{3D}$ criterion \cite{hun88} defines a region as vortex when
\begin{equation}\label{}
	Q_{3D} \equiv \frac{1}{2} (\| \Omega\|^2 - \|S\|^2 ) > 0,
\end{equation}
where $\| A \| = \sqrt{Tr (AA^T)}$ represents the Euclidean norm of the tensor $A$. The two-dimensional equivalent is  based on the two-dimensional horizontal velocity field perpendicular to $\hat{\textbf{z}}$, i.e. on ${\bf \nabla u} |_{2D} = \partial_i u_j (i,j \in {1,2})$, which defines a region as vortex when \cite{vor98,vor02}
\begin{equation}\label{}
	Q_{2D} \equiv 4 \mbox{Det(} {\bf \nabla u}|_{2D}) - [Tr ({\bf \nabla u} |_{2D})]^2 > 0.
\end{equation}
The $Q_{2D}$ criterion is used by Stevens {\it et al.}\ \cite{ste09,ste11b} and Weiss {\it et al.}\ \cite{wei10} to determine the vortex distribution close to the horizontal plate. This analysis has confirmed that vortices are indeed formed close to the horizontal plates. In addition, the analysis revealed that no vortices are formed close to the sidewalls while their distribution is approximately uniform in the rest of the cell. It was  shown by Kunnen {\it et al.}\ \cite{kun10b} that the $Q_{2D}$ criterion is not suitable to recover the smaller scale vortices in the bulk, where the $Q_{3D}$ criterion is more suitable. In addition, they present statistical data, based on simulations and SPIV measurements, on the number and size of vortices at different locations in the flow. Later Stevens {\it et al.}\ \cite{ste12b} compared the regions that are identified as vortex using the $Q_{3D}$ criterion with the regions of warm (cold) fluid that are found with temperature isosurfaces. They show that the vortex regions found with the $Q_{3D}$ criterion agree well with the regions shown by warm (cold) temperature isosurfaces, see figure \ref{fig:figure9}. The main difference is that smaller vortices in the bulk are not shown by the temperature isosurfaces, because their base is not close to the bottom (top) plate where warm (cold) fluid enters the vortices.

\begin{figure}[!t]
  \centering
  \subfigure{\includegraphics[width=0.23\textwidth]{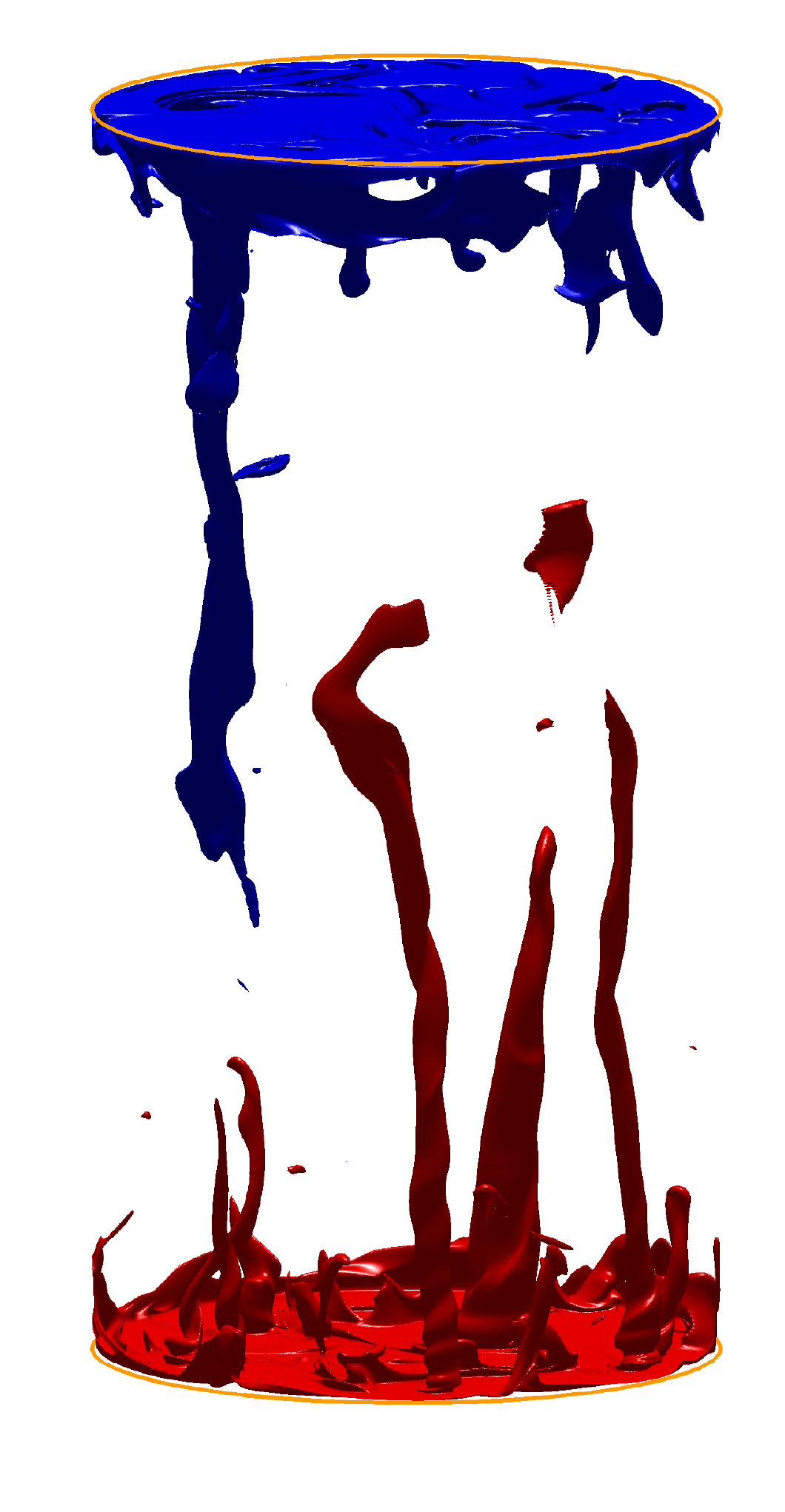}}
  \subfigure{\includegraphics[width=0.23\textwidth]{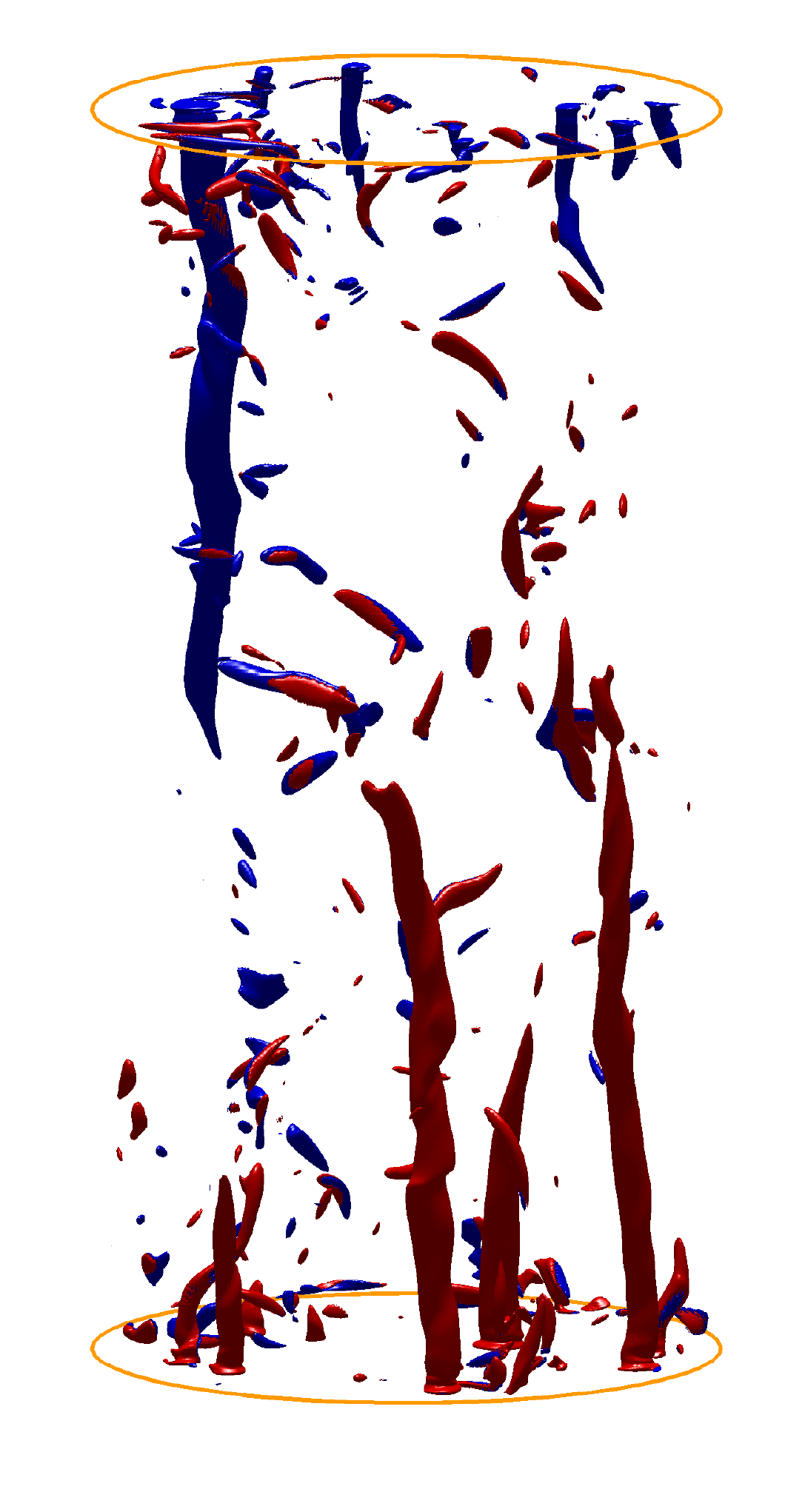}}
  \caption{Flow visualization for $Ra=4.52\times10^9$, $Pr=4.38$, and $1/Ro=3.33$ in a $\Gamma=1/2$ sample. The left image shows a three-dimensional temperature isosurfaces where the red and blue region indicate warm and cold fluid, respectively. The right image shows a visualization of the  vortices, based on the $Q_{3D}$ criterion. Note that the red (blue) regions indicated by the temperature isosurface correspond to vortex regions. Figure taken from Stevens {\it et al.}\ \cite{ste12b}.}
  \label{fig:figure9}
\end{figure}

\section{Influence of the aspect ratio}
In a $\Gamma=1$ sample the onset of heat transport enhancement is visible in temperature measurements at the sidewall by a strong decrease of the relative LSC strength, see figure \ref{fig:figure8}a. However, this is not the case for all aspect ratio samples. Namely, it was revealed by Weiss $\&$ Ahlers \cite{wei11b} that in a $\Gamma=1/2$ sample no strong decrease in the relative LSC strength is observed at the moment that the heat transport enhancement sets in, see figure \ref{fig:figure8}b. These authors discuss that the relative LSC strength according to eq.\ (\ref{Eq Relative Strength LSC}) does not allow one to distinguish between a single convection roll and a two-vortex-state, in which one vortex extends vertically from the bottom into the sample interior and brings up warm fluid, while another vortex transports cold fluid downwards. Because a two-vortex-state results in a periodic azimuthal temperature variation close to the sidewall which cannot be distinguished from the temperature signature of a convection roll with up-flow and down-flow near the side wall. The consequence is that various other criteria, which are based on identifying a cosine variation of the temperature close to the sidewall, also do not show that a transition takes place. Stevens {\it et al.}\ \cite{ste12b} considered this $\Gamma=1/2$ case in numerical simulations and they showed with flow visualization, vortex detection, and the analysis of several statistical quantities that in a $\Gamma=1/2$ sample indeed a flow state that on average can be considered as a two-vortex state is formed. 

The aspect ratio of the sample does not only determine the number of vortices that is found, but also the rotation rate at which heat transport enhancement sets in, see figure \ref{fig:figure10}. Weiss {\it et al.}\ \cite{wei10,wei11} showed with a  Ginzburg-Landau-like model that that the critical rotation rate 1/Ro$_c$ at which the enhancement of the heat transport sets in increases with decreasing aspect ratio due to finite size effects. Based on experimental and numerical data they find that heat transport enhancement sets at at critical Rossby number $Ro_c$
\begin{equation}
\frac{1}{Ro_c} = \frac{a}{\Gamma} \left(1 + \frac{b}{\Gamma}  \right),
\end{equation}
where $a=0.381$ and $b=0.061$. In addition, predictions of the theory about the horizontal vortex distribution in the region close to the sidewall \cite{wei10,wei11} are closely reproduced by numerical measurements \cite{wei10,ste11b}. Although the aspect ratio is important for the heat transport at relatively weak rotation it is shown by Stevens {\it et al.}\ \cite{ste11b} that the aspect ratio dependence of the heat transport disappears for sufficiently strong rotation rates. 

\begin{figure}[!t]
\centering
\subfigure{\includegraphics[width=0.49\textwidth]{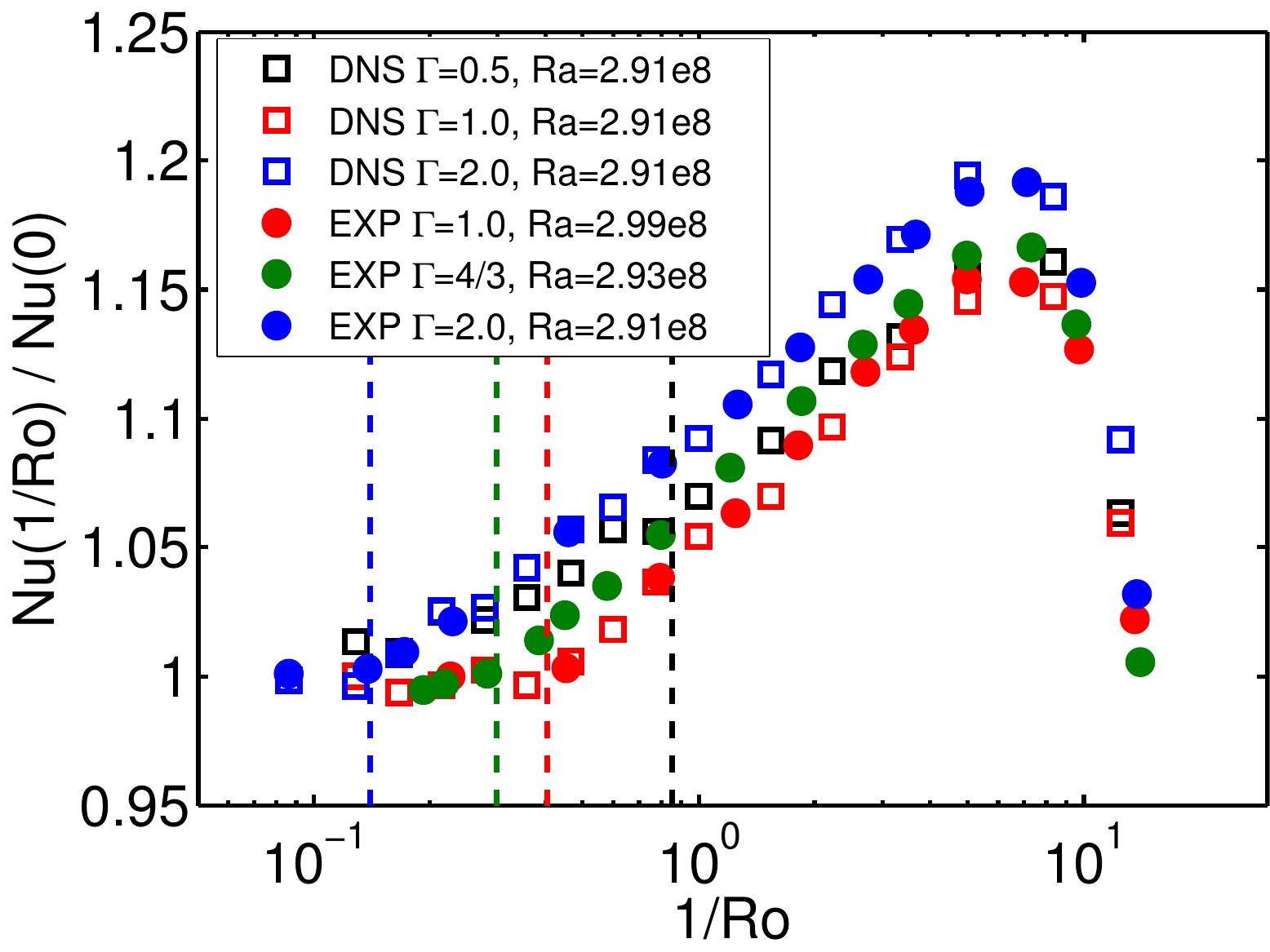}}
\caption{The ratio $Nu(1/Ro)/Nu(0)$ as function of $1/Ro$ for $Ra\approx3\times10^8$ and different $\Gamma$. The experimental results for $\Gamma=1$, $\Gamma=4/3$, and $\Gamma=2$ are indicated in red, dark green, and blue solid circles, respectively. DNS results for $\Gamma=0.5$, $\Gamma=1$, and $\Gamma=2$ are indicated by black, red and blue open squares, respectively. All presented data are for $Pr=4.38$. Figure taken from Stevens {\it et al.}\ \cite{ste11b}.}
\label{fig:figure10}
\end{figure}

\section{Conclusions}
We have summarized and discussed recent works on rotating Rayleigh-Benard (RB) convection and discussed some of our work in more detail. We have seen that a combination of experimental, numerical, and theoretical work has greatly increased our understanding of this problem. As is shown in the rotating RB parameter diagrams, see figure \ref{fig:figure1}, some parts of the parameter space are still relatively unexplored. We especially note that up to now the rotating high $Ra$ number regime has only been achieved in the experiments of Niemela {\it et al.}\ \cite{nie10}. Many natural phenomena like the convection in the atmosphere \cite{har01} and oceans \cite{mar99,rah00} are influenced by rotation. In all these cases the Ra number is very high, and it is thus very important to understand this regime better. It is especially important to know how rotation influences the different turbulent states that are observed in the high $Ra$ number regime \cite{he11}. In addition, it would be very nice to have more detailed data for high and low $Pr$ numbers over a much larger $Ra$ number range, as nowadays most datasets are clustered around $Pr=4.3$ and $Pr=0.7$. Furthermore, there are still several issues that need a deeper understanding. Here we mention the relatively strong  changes in the LSC characteristics that are observed in regime I and the small decrease in the Nusselt number observed in high $Ra$ number experiments just before the onset of heat transport enhancement. It would also be very interesting to get more detailed information about the statistical behavior of the vortical structures in regime II and III and to see whether it is possible to describe all heat transfer properties in rotating RB convection by a unifying model like in the Grossmann-Lohse theory \cite{ahl09} for non-rotating RB convection. 

{\it Acknowledgement:} We benefitted form numerous stimulating discussions with Guenter Ahlers, GertJan van Heijst, Rudie Kunnen, Jim Overkamp, Roberto Verzicco, Stephan Weiss, and Jin-Qiang Zhong over the last years. RJAMS was financially supported by the  Foundation for Fundamental Research on Matter (FOM), which is part of NWO.






\begin{thebibliography}{83}
\providecommand{\natexlab}[1]{#1}
\providecommand{\url}[1]{\texttt{#1}}
\providecommand{\urlprefix}{URL }
\expandafter\ifx\csname urlstyle\endcsname\relax
  \providecommand{\doi}[1]{doi:\discretionary{}{}{}#1}\else
  \providecommand{\doi}[1]{doi:\discretionary{}{}{}\begingroup
  \urlstyle{rm}\url{#1}\endgroup}\fi
\providecommand{\bibinfo}[2]{#2}

\bibitem[{Ahlers et~al.(2009)Ahlers, Grossmann, and Lohse}]{ahl09}
\bibinfo{author}{G.~Ahlers}, \bibinfo{author}{S.~Grossmann},
  \bibinfo{author}{D.~Lohse}, \bibinfo{title}{Heat transfer and large scale
  dynamics in turbulent {{Rayleigh-B\'enard}} convection},
  \bibinfo{journal}{Rev. Mod. Phys.} \bibinfo{volume}{81}
  (\bibinfo{year}{2009}) \bibinfo{pages}{503}.

\bibitem[{Lohse and Xia(2010)}]{loh10}
\bibinfo{author}{D.~Lohse}, \bibinfo{author}{K.~Q. Xia},
  \bibinfo{title}{Small-Scale Properties of Turbulent {{Rayleigh-B\'enard}}
  Convection}, \bibinfo{journal}{Annu. Rev. Fluid Mech.} \bibinfo{volume}{42}
  (\bibinfo{year}{2010}) \bibinfo{pages}{335--364}.

\bibitem[{Johnston(1998)}]{joh98}
\bibinfo{author}{J.~P. Johnston}, \bibinfo{title}{Effects of system rotation on
  turbulence structures: a review relevant to turbomachinery flows},
  \bibinfo{journal}{Int. J. Rot. Mach.} \bibinfo{volume}{4}
  (\bibinfo{year}{1998}) \bibinfo{pages}{97--112}.

\bibitem[{Hartmann et~al.(2001)Hartmann, Moy, and Fu}]{har01}
\bibinfo{author}{D.~L. Hartmann}, \bibinfo{author}{L.~A. Moy},
  \bibinfo{author}{Q.~Fu}, \bibinfo{title}{Tropical convection and the energy
  balance at the top of the atmosphere}, \bibinfo{journal}{J. Climate}
  \bibinfo{volume}{14} (\bibinfo{year}{2001}) \bibinfo{pages}{4495--4511}.

\bibitem[{Marshall and Schott(1999)}]{mar99}
\bibinfo{author}{J.~Marshall}, \bibinfo{author}{F.~Schott},
  \bibinfo{title}{Open-ocean convection: Observations, theory, and models},
  \bibinfo{journal}{Rev. Geophys.} \bibinfo{volume}{37} (\bibinfo{year}{1999})
  \bibinfo{pages}{1--64}.

\bibitem[{Rahmstorf(2000)}]{rah00}
\bibinfo{author}{S.~Rahmstorf}, \bibinfo{title}{The thermohaline ocean
  circulation: A system with dangerous thresholds?}, \bibinfo{journal}{Climate
  Change} \bibinfo{volume}{46} (\bibinfo{year}{2000})
  \bibinfo{pages}{247--256}.

\bibitem[{Hassler et~al.(1999)Hassler, Dammasch, Lemaire, Brekke, Curdt, Mason,
  Vial, and Wilhelm}]{has99}
\bibinfo{author}{D.~Hassler}, \bibinfo{author}{I.~Dammasch},
  \bibinfo{author}{P.~Lemaire}, \bibinfo{author}{P.~Brekke},
  \bibinfo{author}{W.~Curdt}, \bibinfo{author}{H.~Mason},
  \bibinfo{author}{J.-C. Vial}, \bibinfo{author}{K.~Wilhelm},
  \bibinfo{title}{Solar Wind Outflow and the Chromospheric Magnetic Network},
  \bibinfo{journal}{Science} \bibinfo{volume}{283}~(\bibinfo{number}{5403})
  (\bibinfo{year}{1999}) \bibinfo{pages}{810--813}.

\bibitem[{Lappa(2012)}]{lap12}
\bibinfo{author}{Lappa}, \bibinfo{title}{Rayleigh-B\'enard convection with
  rotation}, \bibinfo{publisher}{John Wiley $\&$ Sons, Ltd},
  \bibinfo{address}{Cambridge}, \bibinfo{year}{2012}.

\bibitem[{Stevens et~al.(2011{\natexlab{a}})Stevens, Overkamp, Lohse, and
  Clercx}]{ste11b}
\bibinfo{author}{R.~J. A.~M. Stevens}, \bibinfo{author}{J.~Overkamp},
  \bibinfo{author}{D.~Lohse}, \bibinfo{author}{H.~J.~H. Clercx},
  \bibinfo{title}{Effect of aspect-ratio on vortex distribution and heat
  transfer in rotating {{Rayleigh-B\'enard}}}, \bibinfo{journal}{Phys. Rev. E}
  \bibinfo{volume}{84} (\bibinfo{year}{2011}{\natexlab{a}})
  \bibinfo{pages}{056313}.

\bibitem[{Kunnen et~al.(2008{\natexlab{a}})Kunnen, Clercx, and Geurts}]{kun08e}
\bibinfo{author}{R.~P.~J. Kunnen}, \bibinfo{author}{H.~J.~H. Clercx},
  \bibinfo{author}{B.~J. Geurts}, \bibinfo{title}{Enhanced vertical
  inhomogeneity in turbulent rotating convection}, \bibinfo{journal}{Phys. Rev.
  Lett.} \bibinfo{volume}{101} (\bibinfo{year}{2008}{\natexlab{a}})
  \bibinfo{pages}{174501}.

\bibitem[{Kunnen et~al.(2011)Kunnen, Stevens, Overkamp, Sun, van Heijst, and
  Clercx}]{kun11}
\bibinfo{author}{R.~P.~J. Kunnen}, \bibinfo{author}{R.~J. A.~M. Stevens},
  \bibinfo{author}{J.~Overkamp}, \bibinfo{author}{C.~Sun},
  \bibinfo{author}{G.~J.~F. van Heijst}, \bibinfo{author}{H.~J.~H. Clercx},
  \bibinfo{title}{The role of {{Stewartson}} and {{Ekman}} layers in turbulent
  rotating {{Rayleigh-B\'enard}} convection}, \bibinfo{journal}{J. Fluid.
  Mech.} \bibinfo{volume}{688} (\bibinfo{year}{2011})
  \bibinfo{pages}{422--442}.

\bibitem[{Zhong and Ahlers(2010)}]{zho10c}
\bibinfo{author}{J.-Q. Zhong}, \bibinfo{author}{G.~Ahlers},
  \bibinfo{title}{Heat transport and the large-scale circulation in rotating
  turbulent {{Rayleigh-B\'enard}} convection}, \bibinfo{journal}{J. Fluid
  Mech.} \bibinfo{volume}{665} (\bibinfo{year}{2010})
  \bibinfo{pages}{300--333}.

\bibitem[{Weiss and Ahlers(2011{\natexlab{a}})}]{wei11}
\bibinfo{author}{S.~Weiss}, \bibinfo{author}{G.~Ahlers}, \bibinfo{title}{Heat
  transport by turbulent rotating {{Rayleigh-B\'enard}} convection and its
  dependence on the aspect ratio}, \bibinfo{journal}{J. Fluid. Mech.}
  \bibinfo{volume}{684} (\bibinfo{year}{2011}{\natexlab{a}})
  \bibinfo{pages}{407--426}.

\bibitem[{Hart(2000)}]{har00b}
\bibinfo{author}{J.~E. Hart}, \bibinfo{title}{On the influence of centrifugal
  buoyancy on rotating convection}, \bibinfo{journal}{J. Fluid Mech.}
  \bibinfo{volume}{403} (\bibinfo{year}{2000}) \bibinfo{pages}{133--151}.

\bibitem[{Zhong et~al.(2009)Zhong, Stevens, Clercx, Verzicco, Lohse, and
  Ahlers}]{zho09b}
\bibinfo{author}{J.-Q. Zhong}, \bibinfo{author}{R.~J. A.~M. Stevens},
  \bibinfo{author}{H.~J.~H. Clercx}, \bibinfo{author}{R.~Verzicco},
  \bibinfo{author}{D.~Lohse}, \bibinfo{author}{G.~Ahlers},
  \bibinfo{title}{{{Prandtl}}-, {{Rayleigh}}-, and {{Rossby}}-number dependence
  of heat transport in turbulent rotating {{Rayleigh-B\'enard}} convection},
  \bibinfo{journal}{Phys. Rev. Lett.} \bibinfo{volume}{102}
  (\bibinfo{year}{2009}) \bibinfo{pages}{044502}.

\bibitem[{Stevens et~al.(2010{\natexlab{a}})Stevens, Clercx, and
  Lohse}]{ste10b}
\bibinfo{author}{R.~J. A.~M. Stevens}, \bibinfo{author}{H.~J.~H. Clercx},
  \bibinfo{author}{D.~Lohse}, \bibinfo{title}{Boundary layers in rotating
  weakly turbulent {{Rayleigh-B\'enard}} convection.}, \bibinfo{journal}{Phys.
  Fluids} \bibinfo{volume}{22} (\bibinfo{year}{2010}{\natexlab{a}})
  \bibinfo{pages}{085103}.

\bibitem[{Stevens et~al.(2009)Stevens, Zhong, Clercx, Ahlers, and
  Lohse}]{ste09}
\bibinfo{author}{R.~J. A.~M. Stevens}, \bibinfo{author}{J.-Q. Zhong},
  \bibinfo{author}{H.~J.~H. Clercx}, \bibinfo{author}{G.~Ahlers},
  \bibinfo{author}{D.~Lohse}, \bibinfo{title}{Transitions between turbulent
  states in rotating {{Rayleigh-B\'enard}} convection}, \bibinfo{journal}{Phys.
  Rev. Lett.} \bibinfo{volume}{103} (\bibinfo{year}{2009})
  \bibinfo{pages}{024503}.

\bibitem[{Weiss et~al.(2010)Weiss, Stevens, Zhong, Clercx, Lohse, and
  Ahlers}]{wei10}
\bibinfo{author}{S.~Weiss}, \bibinfo{author}{R.~J. A.~M. Stevens},
  \bibinfo{author}{J.-Q. Zhong}, \bibinfo{author}{H.~J.~H. Clercx},
  \bibinfo{author}{D.~Lohse}, \bibinfo{author}{G.~Ahlers},
  \bibinfo{title}{Finite-size effects lead to supercritical bifurcations in
  turbulent rotating {{Rayleigh-B\'enard}} convection}, \bibinfo{journal}{Phys.
  Rev. Lett.} \bibinfo{volume}{105} (\bibinfo{year}{2010})
  \bibinfo{pages}{224501}.

\bibitem[{Stevens et~al.(2010{\natexlab{b}})Stevens, Clercx, and
  Lohse}]{ste10a}
\bibinfo{author}{R.~J. A.~M. Stevens}, \bibinfo{author}{H.~J.~H. Clercx},
  \bibinfo{author}{D.~Lohse}, \bibinfo{title}{Optimal {{Prandtl}} number for
  heat transfer in rotating {{Rayleigh-B\'enard}} convection},
  \bibinfo{journal}{New J. Phys.} \bibinfo{volume}{12}
  (\bibinfo{year}{2010}{\natexlab{b}}) \bibinfo{pages}{075005}.

\bibitem[{Weiss and Ahlers(2011{\natexlab{b}})}]{wei11b}
\bibinfo{author}{S.~Weiss}, \bibinfo{author}{G.~Ahlers}, \bibinfo{title}{The
  large-scale flow structure in turbulent rotating {{Rayleigh-B\'enard}}
  convection}, \bibinfo{journal}{J. Fluid. Mech.} \bibinfo{volume}{688}
  (\bibinfo{year}{2011}{\natexlab{b}}) \bibinfo{pages}{461--492}.

\bibitem[{Niemela et~al.(2010)Niemela, Babuin, and Sreenivasan}]{nie10}
\bibinfo{author}{J.~Niemela}, \bibinfo{author}{S.~Babuin},
  \bibinfo{author}{K.~Sreenivasan}, \bibinfo{title}{Turbulent rotating
  convection at high Rayleigh and Taylor numbers}, \bibinfo{journal}{J. Fluid
  Mech.} \bibinfo{volume}{649} (\bibinfo{year}{2010}) \bibinfo{pages}{509}.

\bibitem[{Schmitz and Tilgner(2009)}]{sch09}
\bibinfo{author}{S.~Schmitz}, \bibinfo{author}{A.~Tilgner},
  \bibinfo{title}{Heat transport in rotating convection without {{Ekman}}
  layers}, \bibinfo{journal}{Phys. Rev. E} \bibinfo{volume}{80}
  (\bibinfo{year}{2009}) \bibinfo{pages}{015305}.

\bibitem[{King et~al.(2009)King, Stellmach, Noir, Hansen, and Aurnou}]{kin09}
\bibinfo{author}{E.~M. King}, \bibinfo{author}{S.~Stellmach},
  \bibinfo{author}{J.~Noir}, \bibinfo{author}{U.~Hansen},
  \bibinfo{author}{J.~M. Aurnou}, \bibinfo{title}{Boundary layer control of
  rotating convection systems}, \bibinfo{journal}{Nature} \bibinfo{volume}{457}
  (\bibinfo{year}{2009}) \bibinfo{pages}{301}.

\bibitem[{Liu and Ecke(2009)}]{liu09}
\bibinfo{author}{Y.~Liu}, \bibinfo{author}{R.~E. Ecke}, \bibinfo{title}{Heat
  transport measurements in turbulent rotating {{Rayleigh-B\'enard}}
  convection}, \bibinfo{journal}{Phys. Rev. E} \bibinfo{volume}{80}
  (\bibinfo{year}{2009}) \bibinfo{pages}{036314}.

\bibitem[{Liu and Ecke(1997)}]{liu97}
\bibinfo{author}{Y.~Liu}, \bibinfo{author}{R.~E. Ecke}, \bibinfo{title}{Heat
  transport scaling in turbulent {{Rayleigh-B\'enard}} convection: effects of
  rotation and {{Prandtl}} number}, \bibinfo{journal}{Phys. Rev. Lett.}
  \bibinfo{volume}{79} (\bibinfo{year}{1997}) \bibinfo{pages}{2257--2260}.

\bibitem[{Kunnen et~al.(2008{\natexlab{b}})Kunnen, Clercx, and Geurts}]{kun08b}
\bibinfo{author}{R.~P.~J. Kunnen}, \bibinfo{author}{H.~J.~H. Clercx},
  \bibinfo{author}{B.~J. Geurts}, \bibinfo{title}{Breakdown of large-scale
  circulation in turbulent rotating convection}, \bibinfo{journal}{Europhys.
  Lett.} \bibinfo{volume}{84} (\bibinfo{year}{2008}{\natexlab{b}})
  \bibinfo{pages}{24001}.

\bibitem[{Oresta et~al.(2007)Oresta, Stingano, and Verzicco}]{ore07}
\bibinfo{author}{P.~Oresta}, \bibinfo{author}{G.~Stingano},
  \bibinfo{author}{R.~Verzicco}, \bibinfo{title}{Transitional regimes and
  rotation effects in {{Rayleigh-B\'enard}} convection in a slender cylindrical
  cell}, \bibinfo{journal}{Eur. J. Mech.} \bibinfo{volume}{26}
  (\bibinfo{year}{2007}) \bibinfo{pages}{1--14}.

\bibitem[{Kunnen et~al.(2006)Kunnen, Clercx, and Geurts}]{kun06}
\bibinfo{author}{R.~P.~J. Kunnen}, \bibinfo{author}{H.~J.~H. Clercx},
  \bibinfo{author}{B.~J. Geurts}, \bibinfo{title}{Heat flux intensification by
  vortical flow localization in rotating convection}, \bibinfo{journal}{Phys.
  Rev. E} \bibinfo{volume}{74} (\bibinfo{year}{2006}) \bibinfo{pages}{056306}.

\bibitem[{Julien et~al.(1996{\natexlab{a}})Julien, Legg, McWilliams, and
  Werne}]{jul96}
\bibinfo{author}{K.~Julien}, \bibinfo{author}{S.~Legg},
  \bibinfo{author}{J.~McWilliams}, \bibinfo{author}{J.~Werne},
  \bibinfo{title}{Rapidly rotating {{Rayleigh-B\'enard}} convection},
  \bibinfo{journal}{J. Fluid Mech.} \bibinfo{volume}{322}
  (\bibinfo{year}{1996}{\natexlab{a}}) \bibinfo{pages}{243--273}.

\bibitem[{Chandrasekhar(1981)}]{cha81}
\bibinfo{author}{S.~Chandrasekhar}, \bibinfo{title}{Hydrodynamic and
  Hydromagnetic Stability}, \bibinfo{publisher}{Dover}, \bibinfo{address}{New
  York}, \bibinfo{year}{1981}.

\bibitem[{Nakagawa and Frenzen(1955)}]{nak55}
\bibinfo{author}{Y.~Nakagawa}, \bibinfo{author}{P.~Frenzen}, \bibinfo{title}{A
  theoretical and experimental study of cellular convection in rotating
  fluids}, \bibinfo{journal}{Tellus} \bibinfo{volume}{7}~(\bibinfo{number}{1})
  (\bibinfo{year}{1955}) \bibinfo{pages}{1--21}.

\bibitem[{Lucas et~al.(1983)Lucas, Pfotenhauer, and Donnelly}]{luc83}
\bibinfo{author}{P.~J. Lucas}, \bibinfo{author}{J.~Pfotenhauer},
  \bibinfo{author}{R.~Donnelly}, \bibinfo{title}{Stability and heat transfer of
  rotating cryogens. Part 1. Influence of rotation on the onset of convection
  in liquid $^{4}$He}, \bibinfo{journal}{J. Fluid Mech.} \bibinfo{volume}{129}
  (\bibinfo{year}{1983}) \bibinfo{pages}{254--264}.

\bibitem[{Pfotenhauer et~al.(1984)Pfotenhauer, Lucas, and Donnelly}]{pfo84}
\bibinfo{author}{J.~M. Pfotenhauer}, \bibinfo{author}{P.~G.~J. Lucas},
  \bibinfo{author}{R.~J. Donnelly}, \bibinfo{title}{Stability and heat transfer
  of rotating cryogens. Part 2. Effects of rotation on heat-transfer properties
  of convection in liquid {{He}}}, \bibinfo{journal}{J. Fluid Mech.}
  \bibinfo{volume}{145} (\bibinfo{year}{1984}) \bibinfo{pages}{239--252}.

\bibitem[{Pfotenhauer et~al.(1987)Pfotenhauer, Niemela, and Donnelly}]{pfo87}
\bibinfo{author}{J.~Pfotenhauer}, \bibinfo{author}{J.~Niemela},
  \bibinfo{author}{R.~Donnelly}, \bibinfo{title}{Stability and heat transfer of
  rotating cryogens. Part 3. Effects of finite cylindrical geometry and
  rotation on the onset of convection}, \bibinfo{journal}{J. Fluid Mech.}
  \bibinfo{volume}{175} (\bibinfo{year}{1987}) \bibinfo{pages}{85--96}.

\bibitem[{Zhong et~al.(1991)Zhong, Ecke, and Steinberg}]{zho91}
\bibinfo{author}{F.~Zhong}, \bibinfo{author}{R.~E. Ecke},
  \bibinfo{author}{V.~Steinberg}, \bibinfo{title}{Asymmetric modes and the
  transition to vortex structures in rotating {R}ayleigh--{B}{\'e}nard
  convection}, \bibinfo{journal}{Phys. Rev. Lett.} \bibinfo{volume}{67}
  (\bibinfo{year}{1991}) \bibinfo{pages}{2473--2476}.

\bibitem[{Zhong et~al.(1993)Zhong, Ecke, and Steinberg}]{zho93}
\bibinfo{author}{F.~Zhong}, \bibinfo{author}{R.~E. Ecke},
  \bibinfo{author}{V.~Steinberg}, \bibinfo{title}{Rotating
  {{Rayleigh-B\'enard}} convection: asymmetrix modes and vortex states},
  \bibinfo{journal}{J. Fluid Mech.} \bibinfo{volume}{249}
  (\bibinfo{year}{1993}) \bibinfo{pages}{135--159}.

\bibitem[{Hu et~al.(1995)Hu, Ecke, and Ahlers}]{hu95}
\bibinfo{author}{Y.~Hu}, \bibinfo{author}{R.~Ecke},
  \bibinfo{author}{G.~Ahlers}, \bibinfo{title}{Time and Length Scales in
  Rotating Rayleigh-B\'enard Convection}, \bibinfo{journal}{Phys. Rev. Lett.}
  \bibinfo{volume}{74} (\bibinfo{year}{1995}) \bibinfo{pages}{5040--5043}.

\bibitem[{Bajaj et~al.(1998)Bajaj, Liu, Naberhuis, and Ahlers}]{baj98}
\bibinfo{author}{K.~Bajaj}, \bibinfo{author}{J.~Liu},
  \bibinfo{author}{B.~Naberhuis}, \bibinfo{author}{G.~Ahlers},
  \bibinfo{title}{Square Patterns in {{Rayleigh-B\'enard}} Convection with
  Rotation about a Vertical Axis}, \bibinfo{journal}{Phys. Rev. Lett.}
  \bibinfo{volume}{81}~(\bibinfo{number}{4}) (\bibinfo{year}{1998})
  \bibinfo{pages}{806--809}.

\bibitem[{Tagare et~al.(2008)Tagare, Babu, and Rameshwar}]{tag08}
\bibinfo{author}{S.~Tagare}, \bibinfo{author}{A.~B. Babu},
  \bibinfo{author}{Y.~Rameshwar}, \bibinfo{title}{Rayleigh-B\'enard convection
  in rotating fluids}, \bibinfo{journal}{International Journal of Heat and Mass
  Transfer} \bibinfo{volume}{51} (\bibinfo{year}{2008})
  \bibinfo{pages}{1168--1178}.

\bibitem[{Bordja et~al.(2010)Bordja, Tuckerman, Witkowski, Navarro, Barkley,
  and Bessaih}]{bor10}
\bibinfo{author}{L.~Bordja}, \bibinfo{author}{L.~Tuckerman},
  \bibinfo{author}{L.~Witkowski}, \bibinfo{author}{M.~Navarro},
  \bibinfo{author}{D.~Barkley}, \bibinfo{author}{R.~Bessaih},
  \bibinfo{title}{Influence of counter-rotating von K\'arm\'an flow on
  cylindrical Rayleigh-B\'enard convection}, \bibinfo{journal}{Phys. Rev. E}
  \bibinfo{volume}{81} (\bibinfo{year}{2010}) \bibinfo{pages}{036322}.

\bibitem[{Lopez et~al.(2006)Lopez, Rubio, and Marques}]{lop06}
\bibinfo{author}{J.~Lopez}, \bibinfo{author}{A.~Rubio},
  \bibinfo{author}{F.~Marques}, \bibinfo{title}{Travelling circular waves in
  axisymmetric rotating convection}, \bibinfo{journal}{J. Fluid Mech.}
  \bibinfo{volume}{569} (\bibinfo{year}{2006}) \bibinfo{pages}{331--348}.

\bibitem[{Lopez and Marques(2009)}]{lop09}
\bibinfo{author}{J.~Lopez}, \bibinfo{author}{F.~Marques},
  \bibinfo{title}{Centrifugal effects in rotating convection: nonlinear
  dynamics}, \bibinfo{journal}{J. Fluid Mech.} \bibinfo{volume}{628}
  (\bibinfo{year}{2009}) \bibinfo{pages}{269--297}.

\bibitem[{Rubio et~al.(2010)Rubio, Lopez, and Marques}]{rub10}
\bibinfo{author}{A.~Rubio}, \bibinfo{author}{J.~Lopez},
  \bibinfo{author}{F.~Marques}, \bibinfo{title}{Onset of
  {{K\"uppers-Lortz}}-like dynamics in finite rotating thermal convection},
  \bibinfo{journal}{J. Fluid Mech.} \bibinfo{volume}{644}
  (\bibinfo{year}{2010}) \bibinfo{pages}{337357}.

\bibitem[{Scheel and Cross(2005)}]{sch05b}
\bibinfo{author}{J.~D. Scheel}, \bibinfo{author}{M.~C. Cross},
  \bibinfo{title}{Scaling laws for rotating {{Rayleigh-B\'enard}} convection},
  \bibinfo{journal}{Phys. Rev. E} \bibinfo{volume}{72} (\bibinfo{year}{2005})
  \bibinfo{pages}{056315}.

\bibitem[{Scheel et~al.(2010)Scheel, Mutyaba, and Kimmel}]{sch10b}
\bibinfo{author}{J.~D. Scheel}, \bibinfo{author}{P.~L. Mutyaba},
  \bibinfo{author}{T.~Kimmel}, \bibinfo{title}{Patterns in rotating
  {{Rayleigh-B\'enard}} convection at high rotation rates},
  \bibinfo{journal}{J. Fluid Mech.} \bibinfo{volume}{659}
  (\bibinfo{year}{2010}) \bibinfo{pages}{24--42}.

\bibitem[{Rossby(1969)}]{ros69}
\bibinfo{author}{H.~T. Rossby}, \bibinfo{title}{A study of {{B\'enard}}
  convection with and without rotation}, \bibinfo{journal}{J. Fluid Mech.}
  \bibinfo{volume}{36} (\bibinfo{year}{1969}) \bibinfo{pages}{309--335}.

\bibitem[{Julien et~al.(1996{\natexlab{b}})Julien, Legg, McWilliams, and
  Werne}]{jul96b}
\bibinfo{author}{K.~Julien}, \bibinfo{author}{S.~Legg},
  \bibinfo{author}{J.~McWilliams}, \bibinfo{author}{J.~Werne},
  \bibinfo{title}{Hard turbulence in rotating {R}ayleigh--{B}{\'e}nard
  convection}, \bibinfo{journal}{Phys. Rev. E} \bibinfo{volume}{53}
  (\bibinfo{year}{1996}{\natexlab{b}}) \bibinfo{pages}{R5557--R5560}.

\bibitem[{Vorobieff and Ecke(2002)}]{vor02}
\bibinfo{author}{P.~Vorobieff}, \bibinfo{author}{R.~E. Ecke},
  \bibinfo{title}{Turbulent rotating convection: an experimental study},
  \bibinfo{journal}{J. Fluid Mech.} \bibinfo{volume}{458}
  (\bibinfo{year}{2002}) \bibinfo{pages}{191--218}.

\bibitem[{Kunnen et~al.(2010{\natexlab{a}})Kunnen, Geurts, and Clercx}]{kun10}
\bibinfo{author}{R.~P.~J. Kunnen}, \bibinfo{author}{B.~J. Geurts},
  \bibinfo{author}{H.~J.~H. Clercx}, \bibinfo{title}{Experimental and numerical
  investigation of turbulent convection in a rotating cylinder},
  \bibinfo{journal}{J. Fluid Mech.} \bibinfo{volume}{642}
  (\bibinfo{year}{2010}{\natexlab{a}}) \bibinfo{pages}{445--476}.

\bibitem[{Kunnen et~al.(2010{\natexlab{b}})Kunnen, Geurts, and Clercx}]{kun10b}
\bibinfo{author}{R.~P.~J. Kunnen}, \bibinfo{author}{B.~J. Geurts},
  \bibinfo{author}{H.~J.~H. Clercx}, \bibinfo{title}{Vortex statistics in
  turbulent rotating convection}, \bibinfo{journal}{Phys. Rev. E}
  \bibinfo{volume}{82} (\bibinfo{year}{2010}{\natexlab{b}})
  \bibinfo{pages}{036306}.

\bibitem[{King et~al.(2012{\natexlab{a}})King, Stellmach, and Aurnou}]{kin12}
\bibinfo{author}{E.~M. King}, \bibinfo{author}{S.~Stellmach},
  \bibinfo{author}{J.~M. Aurnou}, \bibinfo{title}{Heat transfer by rapidly
  rotating {{Rayleigh-B\'enard}} convection}, \bibinfo{journal}{J. Fluid Mech.}
  \bibinfo{volume}{691} (\bibinfo{year}{2012}{\natexlab{a}})
  \bibinfo{pages}{568--582}.

\bibitem[{Pharasi et~al.(2011)Pharasi, Kannan, Kumar, and
  Bhattacharjee}]{pha11}
\bibinfo{author}{H.~K. Pharasi}, \bibinfo{author}{R.~Kannan},
  \bibinfo{author}{K.~Kumar}, \bibinfo{author}{J.~Bhattacharjee},
  \bibinfo{title}{Turbulence in rotating Rayleigh-B\'enard convection in
  low-Prandtl-number fluids}, \bibinfo{journal}{Phys. Rev. E}
  \bibinfo{volume}{84} (\bibinfo{year}{2011}) \bibinfo{pages}{047301}.

\bibitem[{Raasch and Etling(1991)}]{raa91}
\bibinfo{author}{S.~Raasch}, \bibinfo{author}{E.~Etling},
  \bibinfo{title}{Numerical simulation of rotating turbulent thermal
  convection}, \bibinfo{journal}{Beitr. Phys. Atmosph.} \bibinfo{volume}{64}
  (\bibinfo{year}{1991}) \bibinfo{pages}{185--199}.

\bibitem[{Julien et~al.(1999)Julien, Legg, McWilliams, and Werne}]{jul99}
\bibinfo{author}{K.~Julien}, \bibinfo{author}{S.~Legg},
  \bibinfo{author}{J.~McWilliams}, \bibinfo{author}{J.~Werne},
  \bibinfo{title}{Plumes in rotating convection. Part 1. Ensemble statistics
  and dynamical balances}, \bibinfo{journal}{J. Fluid Mech.}
  \bibinfo{volume}{391} (\bibinfo{year}{1999}) \bibinfo{pages}{151--187}.

\bibitem[{Legg et~al.(2001)Legg, Julien, McWilliams, and Werne}]{leg01}
\bibinfo{author}{S.~Legg}, \bibinfo{author}{K.~Julien},
  \bibinfo{author}{J.~McWilliams}, \bibinfo{author}{J.~Werne},
  \bibinfo{title}{Vertical transport by convection plumes: modification by
  rotation}, \bibinfo{journal}{Phys. Chem. Earth (B)} \bibinfo{volume}{26}
  (\bibinfo{year}{2001}) \bibinfo{pages}{259--262}.

\bibitem[{Husain et~al.(2006)Husain, Baig, and Varshney}]{hus06}
\bibinfo{author}{A.~Husain}, \bibinfo{author}{M.~F. Baig},
  \bibinfo{author}{H.~Varshney}, \bibinfo{title}{Investigation of coherent
  structures in rotating Rayleigh-B\'enard convection}, \bibinfo{journal}{Phys.
  Fluids} \bibinfo{volume}{18} (\bibinfo{year}{2006}) \bibinfo{pages}{125105}.

\bibitem[{Kunnen et~al.(2009)Kunnen, Geurts, and Clercx}]{kun09}
\bibinfo{author}{R.~Kunnen}, \bibinfo{author}{B.~Geurts},
  \bibinfo{author}{H.~Clercx}, \bibinfo{title}{Turbulence statistics and energy
  budget in rotating {{Rayleigh-B\'enard}} convection}, \bibinfo{journal}{Eur.
  J. Mech. B/Fluids} \bibinfo{volume}{28} (\bibinfo{year}{2009})
  \bibinfo{pages}{578 -- 589}.

\bibitem[{Sprague et~al.(2006)Sprague, Julien, Knobloch, and Werne}]{spr06}
\bibinfo{author}{M.~Sprague}, \bibinfo{author}{K.~Julien},
  \bibinfo{author}{E.~Knobloch}, \bibinfo{author}{J.~Werne},
  \bibinfo{title}{Numerical simulation of an asymptotically reduced system for
  rotationally constrained convection}, \bibinfo{journal}{J. Fluid Mech.}
  \bibinfo{volume}{551} (\bibinfo{year}{2006}) \bibinfo{pages}{141--174}.

\bibitem[{Julien et~al.(2012)Julien, Rubio, Grooms, and Knobloch}]{jul12}
\bibinfo{author}{K.~Julien}, \bibinfo{author}{A.~Rubio},
  \bibinfo{author}{I.~Grooms}, \bibinfo{author}{E.~Knobloch},
  \bibinfo{title}{Statistical and physical balances in low Rossby number
  RayleighÐB\'enard convection}, \bibinfo{journal}{Geophysical and
  Astrophysical Fluid Dynamics} \bibinfo{volume}{106} (\bibinfo{year}{2012})
  \bibinfo{pages}{392Ð428}.

\bibitem[{Schmitz and Tilgner(2010)}]{sch10}
\bibinfo{author}{S.~Schmitz}, \bibinfo{author}{A.~Tilgner},
  \bibinfo{title}{Transitions in turbulent rotating {{Rayleigh-B\'enard}}
  convection}, \bibinfo{journal}{Geophysical and Astrophysical Fluid Dynamics}
  \bibinfo{volume}{104} (\bibinfo{year}{2010}) \bibinfo{pages}{481--489}.

\bibitem[{Stevens et~al.(2012)Stevens, Clercx, and Lohse}]{ste12b}
\bibinfo{author}{R.~J. A.~M. Stevens}, \bibinfo{author}{H.~J.~H. Clercx},
  \bibinfo{author}{D.~Lohse}, \bibinfo{title}{Breakdown of the large-scale wind
  in aspect ratio $\Gamma=1/2$ rotating {{Rayleigh-B\'enard}} flow},
  \bibinfo{journal}{Phys. Rev. E} \bibinfo{volume}{86} (\bibinfo{year}{2012}) \bibinfo{pages}{056311}.
  
\bibitem[{Horn et~al.(2011)Horn, Shishkina, and Wagner}]{hor11}
\bibinfo{author}{S.~Horn}, \bibinfo{author}{O.~Shishkina},
  \bibinfo{author}{C.~Wagner}, \bibinfo{title}{The influence of
  {{non-Oberbeck-Boussinesq}} effects on rotating turbulent
  {{Rayleigh-B\'enard}} convection}, \bibinfo{journal}{Journal of Physics:
  Conference Series} \bibinfo{volume}{318} (\bibinfo{year}{2011})
  \bibinfo{pages}{082005}.

\bibitem[{Boubnov and Golitsyn(1990)}]{bou90}
\bibinfo{author}{B.~M. Boubnov}, \bibinfo{author}{G.~S. Golitsyn},
  \bibinfo{title}{Temperature and velocity field regimes of convective motions
  in a rotating plane fluid layer}, \bibinfo{journal}{J. Fluid Mech.}
  \bibinfo{volume}{219} (\bibinfo{year}{1990}) \bibinfo{pages}{215--239}.

\bibitem[{Hart et~al.(2002)Hart, Kittelman, and Ohlsen}]{har02}
\bibinfo{author}{J.~E. Hart}, \bibinfo{author}{S.~Kittelman},
  \bibinfo{author}{D.~R. Ohlsen}, \bibinfo{title}{Mean flow precession and
  temperature probability density functions in turbulent rotating convection},
  \bibinfo{journal}{Phys. Fluids} \bibinfo{volume}{14} (\bibinfo{year}{2002})
  \bibinfo{pages}{955--962}.

\bibitem[{Brown and Ahlers(2006)}]{bro06}
\bibinfo{author}{E.~Brown}, \bibinfo{author}{G.~Ahlers},
  \bibinfo{title}{Rotations and cessations of the large-scale circulation in
  turbulent {{Rayleigh-B\'enard}} convection}, \bibinfo{journal}{J. Fluid
  Mech.} \bibinfo{volume}{568} (\bibinfo{year}{2006})
  \bibinfo{pages}{351--386}.

\bibitem[{Assaf et~al.(2012)Assaf, Angheluta, and Goldenfeld}]{ass12}
\bibinfo{author}{M.~Assaf}, \bibinfo{author}{L.~Angheluta},
  \bibinfo{author}{N.~Goldenfeld}, \bibinfo{title}{Effect of Weak Rotation on
  Large-Scale Circulation Cessations in Turbulent Convection},
  \bibinfo{journal}{Phys. Rev. Lett.} \bibinfo{volume}{109}
  (\bibinfo{year}{2012}) \bibinfo{pages}{074502}.

\bibitem[{Brown and Ahlers(2007)}]{bro07}
\bibinfo{author}{E.~Brown}, \bibinfo{author}{G.~Ahlers},
  \bibinfo{title}{Large-scale circulation model for turbulent
  {{Rayleigh-B\'enard}} convection}, \bibinfo{journal}{Phys. Rev. Lett.}
  \bibinfo{volume}{98} (\bibinfo{year}{2007}) \bibinfo{pages}{134501}.

\bibitem[{Julien and Knobloch(2007)}]{jul07}
\bibinfo{author}{K.~Julien}, \bibinfo{author}{E.~Knobloch},
  \bibinfo{title}{Reduced models for fluid flows with strong constraints},
  \bibinfo{journal}{Journal of Mathematical physics} \bibinfo{volume}{48}
  (\bibinfo{year}{2007}) \bibinfo{pages}{065405}.

\bibitem[{Portegies et~al.(2008)Portegies, Kunnen, van Heijst, and
  Molenaar}]{por08}
\bibinfo{author}{J.~W. Portegies}, \bibinfo{author}{R.~P.~J. Kunnen},
  \bibinfo{author}{G.~J.~F. van Heijst}, \bibinfo{author}{J.~Molenaar},
  \bibinfo{title}{A model for vortical plumes in rotating convection},
  \bibinfo{journal}{Phys. Fluids} \bibinfo{volume}{20} (\bibinfo{year}{2008})
  \bibinfo{pages}{066602}.

\bibitem[{Grooms et~al.(2010)Grooms, Julien, Weiss, and Knobloch}]{gro10}
\bibinfo{author}{I.~Grooms}, \bibinfo{author}{K.~Julien},
  \bibinfo{author}{J.~Weiss}, \bibinfo{author}{E.~Knobloch},
  \bibinfo{title}{Model of Convective {{Taylor}} {{Columns}} in Rotating
  {{Rayleigh}}-{{B\'enard}} Convection}, \bibinfo{journal}{Phys. Rev. Lett.}
  \bibinfo{volume}{104} (\bibinfo{year}{2010}) \bibinfo{pages}{224501}.

\bibitem[{Hart and Olsen(1999)}]{har99}
\bibinfo{author}{J.~E. Hart}, \bibinfo{author}{D.~R. Olsen}, \bibinfo{title}{On
  the thermal offset in turbulent rotating convection}, \bibinfo{journal}{Phys.
  Fluids} \bibinfo{volume}{11} (\bibinfo{year}{1999})
  \bibinfo{pages}{2101--2107}.

\bibitem[{Homsy and Hudson(1969)}]{hom69}
\bibinfo{author}{G.~M. Homsy}, \bibinfo{author}{J.~L. Hudson},
  \bibinfo{title}{Centrifugally driven thermal convection in a rotating
  cylinder}, \bibinfo{journal}{J. Fluid Mech.} \bibinfo{volume}{35}
  (\bibinfo{year}{1969}) \bibinfo{pages}{33--52}.

\bibitem[{Boubnov and Golitsyn(1986)}]{bou86}
\bibinfo{author}{B.~M. Boubnov}, \bibinfo{author}{G.~S. Golitsyn},
  \bibinfo{title}{Experimental study of convective structures in rotating
  fluids}, \bibinfo{journal}{J. Fluid Mech.} \bibinfo{volume}{167}
  (\bibinfo{year}{1986}) \bibinfo{pages}{503--531}.

\bibitem[{Ecke and Liu(1998)}]{eck98}
\bibinfo{author}{R.~E. Ecke}, \bibinfo{author}{Y.~Liu},
  \bibinfo{title}{Traveling-wave and vortex states in rotating
  {R}ayleigh--{B}{\'e}nard convection}, \bibinfo{journal}{Int. J. Eng. Sci.}
  \bibinfo{volume}{36} (\bibinfo{year}{1998}) \bibinfo{pages}{1471--1480}.

\bibitem[{Liu and Ecke(2011)}]{liu11}
\bibinfo{author}{Y.~Liu}, \bibinfo{author}{R.~E. Ecke}, \bibinfo{title}{Local
  temperature measurements in turbulent rotating {{Rayleigh-B\'enard}}
  convection}, \bibinfo{journal}{Phys. Rev. E} \bibinfo{volume}{84}
  (\bibinfo{year}{2011}) \bibinfo{pages}{016311}.

\bibitem[{Stevens et~al.(2011{\natexlab{b}})Stevens, Clercx, and
  Lohse}]{ste10c}
\bibinfo{author}{R.~J. A.~M. Stevens}, \bibinfo{author}{H.~J.~H. Clercx},
  \bibinfo{author}{D.~Lohse}, \bibinfo{title}{Effect of plumes on measuring the
  large-scale circulation in turbulent {{Rayleigh-B\'enard}} convection},
  \bibinfo{journal}{Phys. Fluids} \bibinfo{volume}{23}
  (\bibinfo{year}{2011}{\natexlab{b}}) \bibinfo{pages}{095110}.

\bibitem[{Sakai(1997)}]{sak97}
\bibinfo{author}{S.~Sakai}, \bibinfo{title}{The horizontal scale of rotating
  convection in the geostrophic regime}, \bibinfo{journal}{J. Fluid Mech.}
  \bibinfo{volume}{333} (\bibinfo{year}{1997}) \bibinfo{pages}{85--95}.

\bibitem[{Fernando et~al.(1991)Fernando, Chen, and Boyer}]{fer91}
\bibinfo{author}{H.~J.~S. Fernando}, \bibinfo{author}{R.~Chen},
  \bibinfo{author}{D.~L. Boyer}, \bibinfo{title}{Effects of rotation on
  convective turbulence}, \bibinfo{journal}{J. Fluid Mech.}
  \bibinfo{volume}{228} (\bibinfo{year}{1991}) \bibinfo{pages}{513--547}.

\bibitem[{Vorobieff and Ecke(1998)}]{vor98}
\bibinfo{author}{P.~Vorobieff}, \bibinfo{author}{R.~E. Ecke},
  \bibinfo{title}{Vortex structure in rotating {R}ayleigh--{B}{\'e}nard
  convection}, \bibinfo{journal}{Physica D} \bibinfo{volume}{123}
  (\bibinfo{year}{1998}) \bibinfo{pages}{153--160}.

\bibitem[{King et~al.(2012{\natexlab{b}})King, Stellmach, and Aurnou}]{kin12b}
\bibinfo{author}{E.~M. King}, \bibinfo{author}{S.~Stellmach},
  \bibinfo{author}{J.~M. Aurnou}, \bibinfo{title}{Thermal evidence for Taylor
  columns in turbulent rotating {{Rayleigh-B\'enard}} convection},
  \bibinfo{journal}{Phys. Rev. E} \bibinfo{volume}{85}
  (\bibinfo{year}{2012}{\natexlab{b}}) \bibinfo{pages}{016313}.

\bibitem[{Hunt et~al.(1988)Hunt, Wray, and Moin}]{hun88}
\bibinfo{author}{J.~C.~R. Hunt}, \bibinfo{author}{A.~Wray},
  \bibinfo{author}{P.~Moin}, \bibinfo{title}{Eddies, stream, and convergence
  zones in turbulent flows}, \bibinfo{type}{Report} \bibinfo{number}{CTR-S88},
  \bibinfo{institution}{Center for Turbulence Research}, \bibinfo{year}{1988}.

\bibitem[{Haller(2005)}]{hal05}
\bibinfo{author}{G.~Haller}, \bibinfo{title}{An objective definition of a
  vortex}, \bibinfo{journal}{J. Fluid Mech.} \bibinfo{volume}{525}
  (\bibinfo{year}{2005}) \bibinfo{pages}{1--26}.

\bibitem[{He et~al.(2012)He, Funfschilling, Nobach, Bodenschatz, and
  Ahlers}]{he11}
\bibinfo{author}{X.~He}, \bibinfo{author}{D.~Funfschilling},
  \bibinfo{author}{H.~Nobach}, \bibinfo{author}{E.~Bodenschatz},
  \bibinfo{author}{G.~Ahlers}, \bibinfo{title}{Transition to the ultimate state
  of turbulent {{Rayleigh-B\'enard}} convection}, \bibinfo{journal}{Phys. Rev.
  Lett.} \bibinfo{volume}{108} (\bibinfo{year}{2012}) \bibinfo{pages}{024502}.

\end{thebibliography}







\end{document}